\documentclass[10pt]{article}

\usepackage{amsmath}
\usepackage{amssymb}

\usepackage{graphicx}

\usepackage{cite}

\usepackage{color} 

\usepackage[labelfont=bf,labelsep=period]{caption}

\makeatletter
\renewcommand{\@biblabel}[1]{\quad#1.}
\makeatother

\date{}

\begin{document}

\begin{flushleft}
{\Large
\textbf{Behavioral and Network Origins of Wealth Inequality: Insights from a Virtual World}
}
\\
Benedikt Fuchs$^{1}$, 
Stefan Thurner$^{1,2,3,\ast}$
\\
\bf{1} Section for Science of Complex Systems, Medical University of Vienna, Vienna, Austria
\\
\bf{2} Santa Fe Institute, Santa Fe, New Mexico, USA
\\
\bf{3} International Institute for Applied Systems Analysis, Laxenburg, Austria
\\
$\ast$ E-mail: stefan.thurner@meduniwien.ac.at
\end{flushleft}

\section*{Abstract}
 Almost universally, wealth is not distributed uniformly within societies or economies. 
 Even though wealth data have been collected in various forms for centuries, the origins for the observed wealth-disparity and 
 social inequality are not yet fully understood. 
 Especially the impact and connections of human behavior on wealth could so far not be inferred from data.
 Here we study wealth data from the virtual economy of the massive multiplayer online game (MMOG) \emph{Pardus}.
 This data not only contains every player's wealth at every point in time, but also all actions of every player over a timespan of almost a decade.
 We find that wealth distributions in the virtual world are very similar to those in western countries. In particular we find an approximate 
 exponential for low wealth and a power-law tail.
 The Gini index is found to be $g=0.65$, which is close to the indices of many Western countries.
 We find that wealth-increase rates depend on the time when players entered the game. Players that entered the game early on tend 
 to have remarkably higher wealth-increase rates than those who joined later. 
 Studying the players' positions within their social networks, we find that the local position in the trade network is most relevant for wealth.
 Wealthy people have high in- and out-degree in the trade network, relatively low nearest-neighbor degree and a low clustering coefficient. 
 Wealthy players have many mutual friendships and are socially well respected by others, 
 but spend more time on business than on socializing.
 We find that players that are not organized within social groups with at least three members are significantly poorer on average.  
 We observe that high `political' status and high wealth go hand in hand.
 Wealthy players have few personal enemies, but show animosity towards players that behave as public enemies.


%

\section*{Introduction}
The richest 1\% own nearly half of all global wealth. The richest 10\% claim about 86\% of global wealth, 
so that 90\% of the world's population have to share the rest \cite{CS}.
Wealth is not distributed evenly across nations nor within economies. The inequality of wealth is a strong 
driving force in human history and has been given much attention ever since the onset of economics. 
The definition of wealth is not straight forward, and varies widely across history and schools of thought. 
Adam Smith uses the word \emph{stock} for the personal possessions and regards everything except material 
goods as per se worthless \cite{AdamSmith}. Wealth is defined by Thomas R. Malthus  as ``those material objects which 
are necessary, useful, or agreeable to mankind'' \cite[p. 28]{Malthus}, 
and by John S. Mill as ``all useful or agreeable things which possess exchangeable value'' \cite[p. 10]{Mill1848}. 
Alfred Marshall in his definition includes immaterial goods, such as  personal skills, as long as they can be transferred \cite{Marshall}. 
To accumulate wealth, income must exceed the needs for immediate survival \cite{AdamSmith,Mill1848,Marshall}, which 
implies that a society living at the subsistence level is basically egalitarian,  since no one  can accumulate wealth.
As soon as societies produce a surplus, social stratification arises \cite{Mill1848,Engels1883,Childe1944,Herskovits1952},
and universally leads to an unbalanced distribution of wealth \cite{Herskovits1940}. 

The quantitative study of personal wealth distributions begins with Vilfredo Pareto \cite{Pareto1896},
who observed in a variety of datasets that the tails of wealth distributions follow a power-law, $p(w)\sim w ^{-\alpha}$. 
Pareto thought that this power-law appears universally across times and nations. Indeed it is found in an impressive 
number of data, including ancient Egypt, medieval Hungary, present day Europe, UK,  USA, Russia, India, and China %
\cite{Jayadev2008,Dragulescu2001,Abulmagd2004,Hegyi2007,Klass2007,Sinha2006,Ning2007,Steindl1965}. 
We present a collection of data in Tab. S1 in the SI. 
%
For these countries the power exponents range from $0.7$ to $2.44$. 
Datasets containing both the bulk of the population and the top richest show a double power-law \cite{Coelho2008}:
While exponents dealing with the richest, like \cite{Hegyi2007,Klass2007,Sinha2006,Ning2007}, are close to (sometimes below) 1,
exponents describing the bulk of the population, like \cite{Steindl1965,Jayadev2008,Dragulescu2001} are found to be around 2.
In Pardus, the extremely rich class is absent. 
The power exponent $2.46$ found in Pardus is in the range, but on the high end, of exponents describing the moderately rich.

Empirical data of wealth distributions is a non-trivial issue, the main difficulty being to obtain correct data on the wealth of individuals 
\cite{Marshall, Banerjee2010}.
Most countries have an income tax, only a few employ a wealth tax. Out of the 158 countries and territories listed in \cite{EY2013}, 
149 levy tax on income, and only seven on wealth. 
Income tax data can be used to generate income distributions to study wealth-increase and re-distribution dynamics of the  
low- and medium income classes. 
Sometimes income has been used as a proxy for wealth \cite{Souma2000,Clementi2005,Scafetta2004,Ferrero2005,Clementi2007},
with the problematic assumption that income is approximately proportional to wealth plus human capital \cite{Angle1986}.
Income of the richest is often not reflected in income tax data, since their wealth increments are usually not 
related to salaries, but are usually due to capital gains.
Therefore the tail of the distribution is often not seen in tax-based data: wealth distribution data poses a challenge to this day.

\begin{figure}[!htb]
\begin{center}
 \includegraphics[width=\textwidth]{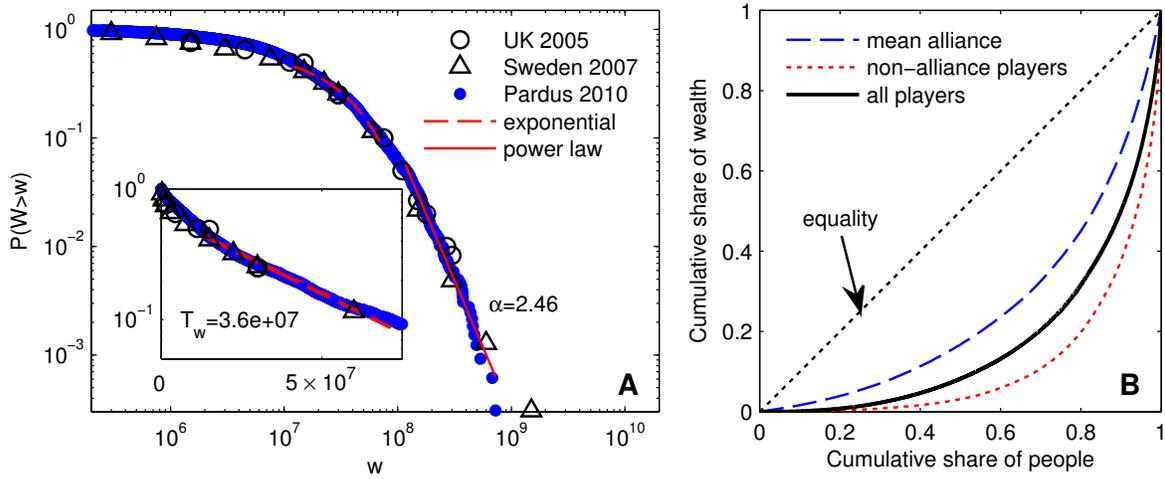}
\end{center}
 \caption{
 {\bf Wealth distribution and Lorenz curve.} 
A Cumulative wealth distributions for the United Kingdom, %
Sweden\protect\footnotemark, and for the %
 Pardus MMOG on day 1200. %
 People with negative wealth have been excluded. %
 A power-law tail is visible. 
 The exponent is determined with a least square fit to the richest 5\% of the population. %
 The bulk of the distribution , i.e. the richest 50\% to 10\%, can be fitted with an exponential function $P(W>w) \propto \exp(w/T_w)$ (inset). %
 The poorest obviously do not follow an exponential distribution, while the richest 10\% are the crossover region to a power-law. %
 B Lorenz curve of wealth in Pardus on day 1200 (excluding newcomers and inactive players). 
 For every alliance, a separate Lorenz curve is calculated. The dashed blue curve represents the average of these %
 single alliance Lorenz curves. %
 }
 \label{fig:Pardus_wealth_cum}
\end{figure}
\footnotetext{Data sources: HM Revenue \& Customs (2005) Personal wealth, distribution among the adult population of%
  marketable wealth (Series C). Available: http://www.hmrc.gov.uk/stats/personal\_wealth/13-5-table-2005.pdf. Accessed 30 January 2014. \\ 
Statistics Sweden (2010) Net wealth in different intervals 2007. number of persons, mean value%
  and sum (corrected 2010-03-22). Available: http://www.scb.se/en\_/Finding-statistics/Statistics-by-subject-area/Household-finances/Income-and-income-distribution/Households-assets-and-debts/Aktuell-Pong/2007A01K/Net-wealth-in-different-intervals-2007-Number-of-persons-mean-value-and-sum-Corrected-2010-03-22/. Accessed 11 February 2014}

In this work we are primarily interested in wealth, rather than income distributions, and attempt for its explanation in terms of behavioral and network aspects. 
Data on wealth distributions is obtainable  from countries imposing a tax on wealth like Sweden\footnote{In 2007 the wealth tax in Sweden was abolished  \cite{EY2013}.},
surveys on wealth \cite{Jayadev2008},
adaptions of data on inheritance tax \cite{Dragulescu2001},
the size of houses found in an excavation \cite{Abulmagd2004},
the number of serfs from a historical almanac \cite{Hegyi2007},
and top-rich rankings in magazines \cite{Sinha2006,Klass2007,Ning2007}.
In Fig. \ref{fig:Pardus_wealth_cum} A the wealth distributions for the UK in 2005 and Sweden in 2007 are shown. 
Both exhibit a power-law tail, whereas the bulk of the distribution is better described with an exponential (inset).  
There is evidence that in many economies the wealth distribution for small wealth levels follows an approximate exponential function  \cite{Dragulescu2001}, 
whereas the tail follows an approximate power-law 
\cite{Pareto1896,Abulmagd2004,Hegyi2007,Jayadev2008,Steindl1965,Dragulescu2001,Klass2007,Sinha2006,Ning2007,Coelho2008}.
Consumption can not sustainably drop below the minimum income needed to exist.
To avoid the consequences of consumption below the minimum income needed to exist, many modern countries provide welfare.
This leads to the situation that a significant fraction of the population can have practically no wealth
(for example 24\% of Swedish households had negative or zero net wealth in 1992 \cite{Domeij2002}), but very few have income below the minimum needed to exist.

A number of models have been suggested to understand the features of empirical wealth distributions and relate them to appropriate mechanisms. 
While power-law distributions can be understood by a multiplicative (e.g. rate of return) re-distribution processes 
that favors the part of the population that are wealthy enough to hold substantial financial assets, the bulk of the distribution can be understood by 
relatively simple exchange models.
The first models that explain a power-law income distribution (in most cases the tail) were brought forward by \cite{Champernowne1953}.
A model incorporating both proportional growth and exchange was suggested in \cite{Bouchaud2000}: $\textstyle{\frac {\mathrm{d}w_i} {\mathrm{d}t}} = \epsilon_i(t) w_i(t) + J\left(\langle w(t) \rangle - w_i(t)\right)$.
Here $w_i(t)$ is the wealth of individual $i$ at time (step) $t$, $J$ is a coupling constant and $\epsilon$ is a random variable with mean 0 and finite standard deviation $\sigma$ independent of $i$ and $t$. 
The model has a stable solution with an asymptotic power-law for large $w$ with a power exponent of $\alpha=1+\textstyle{\frac J {\sigma^2}}$.
In \cite{Silva2005,Banerjee2010}, a Fokker-Planck equation model was proposed for the income distribution.  
It leads to an income distribution that behaves like an exponential for small and mid-range incomes,  
and as a power-law for the highest incomes. The interpolating between the bulk and the rich is different than in \cite{Bouchaud2000}. 
The model has been extended to capture a second power-law for the super-rich \cite{Jagielski2013}.
To understand the exponential distribution of the bulk, simple additive wealth exchange models can be used. 
For example in \cite{Dragulescu2000}
at each time step $t$, a pair of agents $i$ and $j$ is  chosen randomly, and exchange an amount $\Delta w$ of money, 
so that $w_i(t+1)=w_i(t)+\Delta w$, and $w_j(t+1)=w_j(t)-\Delta w$. To avoid agents with infinite debt, a minimum (negative) 
wealth is imposed so that the exchange only takes place if $w_j(t)\geq w_{min} + \Delta w$.
Adding a savings propensity $\lambda$ to the exchange model \cite{Dragulescu2000} means that 
agents use only a fraction $(1-\lambda)$ of their wealth for exchange, $\Delta w = (1-\lambda)\left[ \epsilon(w_i(t) + w_j(t))-w_i(t) \right]$. 
Here $\epsilon$ is a random variable between zero and one.
This leads to a Gamma distribution of wealth $P(W>w) \propto w^\lambda \exp(b w)$ \cite{Chakraborti2000}, with $b$ a constant.
If this savings propensity is drawn from an uniform distribution over $[0, 1]$, a distribution with power-law tail follows \cite{Chatterjee2007}.
Another model that leads to the Gamma distribution is derived from the concept of social stratification. The model is given by  
$w_i(t+1)=w_i(t)+d Zw_j(t)-(1-d)Zw_i(t)$, $w_j(t+1)=w_j(t)+(1-d) Zw_i(t)-dZw_j(t)$, 
where individuals $i$ and $j$ are chosen randomly at each step, 
 $Z \in [0, 1]$ is a random variable,  and $d$ is a binary random variable, zero or one 
\cite{Angle1986}.
The resulting function has been used to fit income distributions of the UK and USA \cite{Scafetta2004}.
%
There are several models of multiplicative wealth growth \cite{Gibrat1931},  $w_i(t+\tau)-w_i(t) = \epsilon_i(t) w_i(t)$,
that lead to log-normal cumulative distributions, $P(W>w) = \frac{\beta}{w\sqrt{\pi}} \exp\left[-(\beta (\ln w - w_0))^2\right]$. 
Models of this kind have been used to describe income distributions \cite{Souma2000, Clementi2005}.
%
%
Other functions that effectively interpolate between an exponential in the low wealth regime and a power-law tail, include
the Tsallis distribution ($q$-exponential), $p(w) = (2-q) \lambda [1-(1-q)\lambda w]^\frac 1 {1-q}$, 
which has been applied to the distribution of income in Japan, UK and New Zealand \cite{Ferrero2005}, with $q\sim1.1$.
Another generalization of the exponential function,
$P(\hat{W}>\hat{w}) = \operatorname{exp_\kappa}(-\beta\hat{w}^\alpha) = \left( \sqrt{1 + \kappa^2 (\beta\hat{w}^\alpha)^2} - \kappa \beta\hat{w}^\alpha \right)^{1/\kappa}$,
with $\hat{w} \equiv w/\langle w \rangle$, $\alpha, \beta>0$, and $\kappa \in [0, 1)$,
has been fitted to income distributions of Germany, Italy, and UK \cite{Clementi2007}.

It was hitherto impossible to directly study wealth of individuals as a consequence of social performance indicators, positions and roles  
within social networks, or behavioral patterns.  However, in the context of massive multiplayer online games (MMOG) there exists 
an opportunity to study the origin of wealth of individuals as a function of their position within their social networks and behavioral patterns. 
In this paper we use data from the MMOG \emph{Pardus}, 
where people live a virtual life in synthetic (computer game) worlds \cite{Castronova2005}.
The essence of MMOGs is the open-ended simultaneous interaction of thousands of players in a multitude of ways, 
including communication, trade, and accumulation of social status. The number of `inhabitants' of some of 
these virtual worlds exceeds the population of small countries: World of Warcraft, started in 2004 and currently the 
biggest MMOG worldwide, has about 7.7 millions of paying subscribers as of June 2013 \cite{Blizzard2013}.
Production and trade between players is  a common feature of many MMOGs, and can create a complex and highly structured 
economy within the game. Although all goods produced and traded are virtual, the economy as such is real:
players invest time and effort to invent, produce, distribute, consume and dispose  these virtual goods and services.
Virtual goods produced in some MMOGs can be traded in the real world for real money, 
which then allows to measure hourly wage and gross national product of a MMOG \cite{Castronova2001}.
In some MMOGs, entire characters (avatars) are traded for money in the real world, which allows to quantify 
`human capital' , such as skills, influence on others, leadership, etc. Economical and sociological data are easily 
accessible in virtual worlds in terms of log-files, and have become a natural field for research %
 \cite{Malaby2006,Bainbridge2007,Szell2010msd,Klimek2013,Szell2010mol,Szell2012,Thurner2012,Szell2013,Guo2012},
even allowing economical experiments \cite{Castronova2008}.

The particular dataset of the Pardus game comprises complete information about a virtual, but nevertheless human, society. 
We have complete knowledge of every action, interaction, communication, trade, location change, etc. of each of the 40,785 players at 
the time resolution of one second. 
The society of the Pardus game has been studied extensively over the past years. 
The social networks have been quantified with respect to their structure and dynamics, 
revealing network densification \cite{Leskovec2007}, corroborating the ``weak ties hypothesis'', and showing evidence for triadic 
closure as driving mechanism for the evolution of the socially positive networks \cite{Szell2010msd,Klimek2013}.
The empirical multiplex nature of the social networks allows to quantify correlations between socially positive interactions, 
and  between various types of interactions \cite{Szell2010mol}.
Mobility of avatars as studied within the Pardus world shows striking similarities to human travel in the real world \cite{Szell2012}.
Timeseries of actions in the Pardus game have been used to quantify the origin of good an cooperative behavior, 
and to predict actions of avatars, given the knowledge of their past actions \cite{Thurner2012}. Social network formation dynamics within 
Pardus have been used to demonstrate the existence of gender differences in the social networking behavior of male and female avatars \cite{Szell2013}.

\subsection*{The MMOG {Pardus}} \label{sec:pardus}

The MMOG {Pardus} provides a persistent synthetic world
in which thousands of players interact through their game characters (avatars) which they control through their browser.
Players tend to identify with their characters \cite{Castronova2005},
which allows us to write ``player'' for ``the player's character'' in the following.
The setting of Pardus is futuristic. Every player owns a space ship to travel the universe, 
which contains planets, space cities,  natural resource fields, and even space monsters.
Players can explore the universe, build production sites (factories) and trade with each other, and fight each other, or monsters.
Many players are driven by the accumulation of `social status' by obtaining honors for certain social achievements of by purchasing 
expensive items that serve as status symbols. There is no overall goal in the game, and players constantly define their own goals and roles.
Pardus is free of charge but requires registration. In total more than 400,000 players have registered since 2004.
Pardus has an internal `unit of time' called \emph{Action Points} (AP). At every day every player has a limited number of 5000 APs that can be spent. 
Different actions of the player `cost' various amounts of APs. 

\subsubsection*{The economy of Pardus}
The input-output production matrix of the economy and the variety of goods are pre-defined within the Pardus framework.
Goods are of completely uniform quality (homogeneous). 
Consumables and equipment can be partially substituted by other types of consumables and equipment,
while intermediate goods are needed for production in exact proportions.
There are five commodities that are natural resources, 19 serve as intermediate goods, and five are end-products, i.e. consumables.

Although capital requirements to create production facilities are low, there are barriers to entering production. 
Incumbents may threaten or harm potential new entrepreneurs. Game rules set a maximum number of 
production facilities for every single player. Many players operate the maximum number of factories.
Production facilities in Pardus are fixed assets with infinite durability but can not be sold. 
Investments into production facilities therefore motivate incumbents to stay in the sector (exit barrier).
While no labor is needed for production itself, transport of raw materials and intermediate goods requires effort and resources 
of the players. Because of transport costs, facilities effectively only compete with similar facilities which are close by.
Together with sparse distribution of production facilities, this leads to effective local oligopolies.

A special kind of goods are various forms of equipment, i.e. items like a space ship or weapons. Equipment can only be produced in special facilities 
which also act as warehouses and selling points. Equipment is durable, but has a finite lifetime. Maintenance applied to equipment 
can increase the lifetime. 
When a player sells equipment after usage, it is scrapped\footnote{Players may own only a limited amount of equipment, 
resulting in an incentive to sell from time to time.}.
Owners of production facilities are completely free to set the price at which they buy their raw materials and sell their products.
There are also non-player facilities (belonging to the game itself) whose prices directly react to local supply and demand within certain limits.
The monetary currency of Pardus is called \emph{credits}.
There is no credit or banking system and all transactions are payed and cleared immediately. 
There is no inflation in the game.

\subsubsection*{The social groups in Pardus}\label{sec:alliances}
Players can organize themselves within  social groups for various purposes. Groups often share
the same interests, or are constituted as pirate groups, exploration teams, self-defense units, etc. 
Usually groups do not get larger than about 140 members.
Pardus provides administration tools for officially declared groups, which are then called \emph{alliances}.
Alliances have a common cash pool which they use for their goals, like defense or production.

Often alliances are created and used for economic purposes. 
Alliances can locally coordinate production capacities to build up entire production chains.
For an optimal production chain, it is sometimes necessary to increase the production capacity of a certain intermediate good.
This is often done by luring a new member into the corresponding business and by paying her for the construction of an additional production facility.

\subsubsection*{Wealth of a player}

There are several ways for players to obtain wealth: 
 trading,
 collecting natural resources,
 producing goods,
 working for hire (most common jobs are courier/teamster, hunter, or bounty hunter),
 receiving donations or other payments,
 an increase of the alliance funds (by payments from someone else),
 and robbing or stealing.

We define wealth or `personal net-worth' of player $i$ as the sum of the value of his assets, 
 i.e. liquidity (cash) $v_{l,i}$,  
 equipment $v_{e,i}$,   
 share of alliance funds $v_{af,i}$,  
 and inventory $v_{inv, i}$.
The latter are all the commodities stored in player $i$'s production facilities and in the space ship.
Equipment are various in-game items like a space ship or weapons.
Each type of equipment can be bought (new) at varying prices and sold (used) at a constant price.
At non-player facilities, equipment can be bought for twice the sell price.
To determine the contribution to the net-worth, we therefore take 1.5 times the sell price as the `value' of each piece of equipment.
The values of the different types of equipment span five orders of magnitude.
The share of the alliance funds, if the player is a member of an official group, is calculated by evenly dividing the group's cash pool to all members.
Additionally, it is discounted by a factor of two.
Inventory is neglected, an exception being those warehouses that are associated with the production of equipment.
Real estate, i.e. the production facilities, can not be sold and therefore has no market value.

There are several ways to reduce wealth in the game:
 consuming,
 paying for maintenance (either because of `natural' degradation or because of damage from a fight),
 investing into production facilities or equipment,
 discarding goods,
 becoming victim of theft or robbery,
 giving to fellow players or paying into the alliance funds,
 a decrease of the alliance funds,
 or making an adverse trade.
In summary, the wealth of individual $i$ at time $t$ is given by:
\begin{equation}
 w_i(t) = v_{l,i}(t) + v_{e,i}(t) + v_{af,i} + v_{inv, i} \quad .
\end{equation}

In the following we use a series of measures that are necessary to quantify wealth and performance of the avatars.
To quantify \emph{wealth} we use $w_i(t)$ for the momentary wealth of player $i$ at time $t$.
The \emph{age} of a player is the number of days since the player entered the game for the first time. 
We measure the cumulative activity of a player by the total amount of APs he has `spent'.
We denote this cumulative total activity by $a_i(t)$.
The \emph{wealth-gain} of player $i$ we denote by
\begin{equation}\label{eq:defEta}
 \eta_i(t)\equiv w_i(t)/a_i(t) \quad ,
\end{equation}
which is measured in credits per AP.
$\eta_i(t)$ can also be seen as efficiency at gaining wealth.

There are a number of achievement-factors in the game that measure certain properties of players.
The efficiency harvesting natural resources is quantified (as a game feature) by the  \emph{farming skill} of a player. 
Other performance related measures that players can gain and lose over time, are the \emph{combat skill} that quantifies fighting skills, 
and the \emph{experience points} (XPs), which keep a record of \emph{fighting experience} and other activities.
Players may choose to be member of a `political' \emph{faction}, which sometimes engage in large-scale conflict (war) against each other.
The \emph{faction rank} is a measure of influence in one's faction: above a certain threshold, it grants the privilege to take part in the decision on war or peace.
It is gained by several specific activities.
Some players regard high combat skill, faction rank, or XP as their main goals in Pardus.

\section*{Results}
\subsection*{The wealth distribution} \label{sec:distribution}

\begin{table}[!ht]
\caption{\bf{Measures of wealth-inequality in Pardus compared to real-world countries}}
 \begin{tabular}{l c l  *{4}{c}}
  country & year & unit & Gini index $g$ & bottom 50\% & top 10\% & top 1\% \\
  \hline
  Pardus & 2010 & all players & 0.653 & 8.2 & 49.9 &  12.4 \\
  Pardus & 2010 & alliance players & 0.495 & 16.7 & 35.4 & 4.6 \\
  Pardus & 2010 & non-alliance players & 0.701 & 3.1 & 62.3 & 20.2 \\
  \hline
  China & 2002 & individual & 0.550 & 14.4 & 41.4 & -- \\
  France & 1994 & adult & 0.730 & -- & 61.0 & 21.3 \\
  Germany & 1998 & household & 0.667 & 3.9 & 44.4 & -- \\
  UK & 2000 & adult & 0.697 & 5.0 & 56.0 & 23.0 \\
  USA & 2001 & family & 0.801 & 2.8 & 69.8 & 32.7 \\
 \end{tabular} 
 \begin{flushleft}Gini index $g$ for wealth, and fraction of total wealth in \% held by a fraction of the population.
  $g$ ranges from 0 for complete equality to 1 for extremest inequality (see Methods). 
  Real-world data is taken from \cite{Davies2011a}. 
 \end{flushleft}
 \label{tab:GiniLorenz}
\end{table}

Figure \ref{fig:Pardus_wealth_cum} A shows the wealth distribution of Pardus in comparison to the UK and Sweden. 
From the Pardus data, new and inactive players have been excluded, see Methods.
The bulk of the (cumulative) distribution is compatible with an exponential \cite{Dragulescu2001} with decay $T_w=3.6\times10^7$ credits.
 The tail is best fitted with a power-law with exponent $\alpha=2.46$. 
For comparison Tab. S1 in the SI contains power exponents for several `real' countries.
Figure \ref{fig:Pardus_wealth_cum} B shows the Lorenz curve (see Methods) for the Pardus society (black line). 
The closer the Lorenz curve is to the diagonal (black dotted line) the more homogenous is the wealth distribution. 
Uniform wealth distribution corresponds to the diagonal.
Associated to this line is a Gini index \cite{Gini1912} of $g=0.653$. For comparison with `real' countries see Tab. \ref{tab:GiniLorenz}.
We further show the Lorenz curve for all players that are not organized in any alliance (red dotted line). These players  generally operate individually, 
and show a much more pronounced wealth inequality than the entire society, the respective Gini index being $0.701$. 
In contrast, the Lorenz curve for the various alliances (the average over all alliances with at least 5 members is shown as a dashed blue line) indicates that people within the alliances tend to be much more equal in wealth, when compared to the entire society.
The Gini index for the alliances is  $0.495$.
The main reason for this higher equality is the smaller fraction of poor players in alliances: 
while $79\%$ of the total population and $92\%$ of the richest $10\%$ are alliance members, only $28\%$ of the poorest $10\%$ are.

\subsubsection*{Evolution of the wealth distribution over time}

\begin{figure}[!htbp]
\begin{center}
 \includegraphics[width=\textwidth]{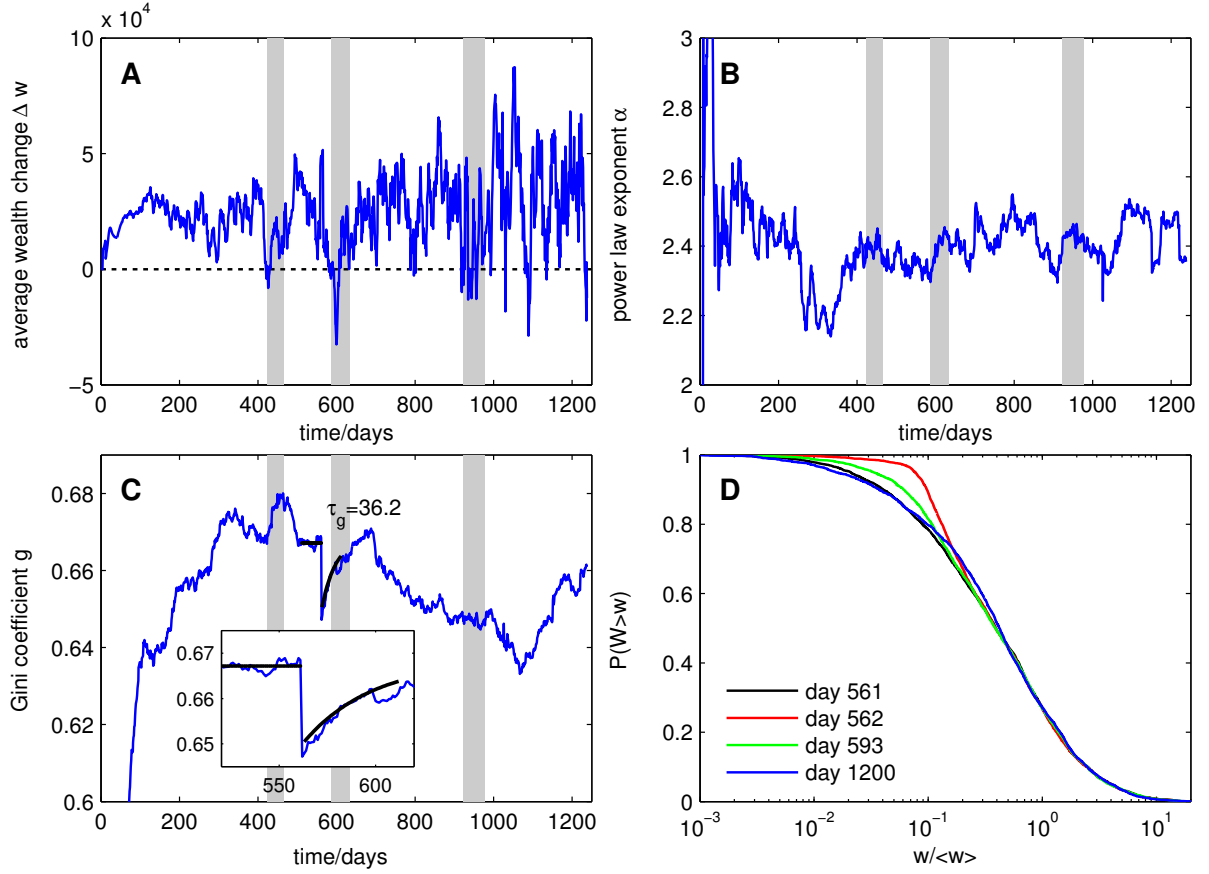}
\end{center}
 \caption{{\bf Time evolution of the wealth distribution in the entire society.}
A Seven day moving average of the change of the average wealth $\Delta w(t)$.
B Power-law exponent $\alpha(t)$, 
C Gini index $g(t)$.
D Scaled wealth distribution at four different days (rescaled by average wealth).
Gray shaded areas indicate periods of large scale war in the game. A `Christmas charity' event on day 562
led to a re-distribution from the wealthy to the poor, resulting in a downward jump of the Gini index.
The inset shows the exponential recovery to the previous level. }
 \label{fig:ginietc_evolution}
\end{figure}

The average wealth in the Pardus society $\langle w(t) \rangle$ grows over time. Brackets indicate the average over all players 
present at time $t$. The daily average change we denote by $\Delta w(t) \equiv \langle w(t) \rangle-\langle w(t-1) \rangle$, and is presented in 
Fig. \ref{fig:ginietc_evolution} A. We find that $\Delta w(t) >0$ on 83\% of all days, and that the average daily increase of 
average wealth is $2.42\times 10^4$ credits.
In Fig. \ref{fig:ginietc_evolution} A it is visible that average wealth increases less during war periods (gray shaded areas):
the average daily increase during the three war periods is $9.74\times 10^3$, $3.22\times 10^3$, and $1.26\times 10^4$ credits respectively, 
against $2.64\times 10^4$ credits in peace times.
Figure \ref{fig:ginietc_evolution} B shows the evolution of the power-law exponent $\alpha$. Its value is limited to a region
between $2.14$ and $2.55$. After an initial steep rise in the first 150 days, the Gini index $g$ fluctuates between a maximum of $0.68$ and a minimum of $0.63$,
as seen in  Fig. \ref{fig:ginietc_evolution} C.
A prominent feature is a sharp drop of $g$ from $0.67$ to $0.65$ on day 562 which corresponds to 2008/12/24.
At this day, a `global' charity event took place, where thousands of players donated cash for the less wealthy.
The inset indicates an exponential recovery, $g(t=561)-g(t-561) \propto \exp (-(t-561)/\tau_g)$ with decay time $\tau_g = 36.2$ days, (black line).
This indicates a remarkable stability of the shape of the wealth distribution, as also seen in Fig. \ref{fig:ginietc_evolution} D:
First, after dividing wealth by the average wealth on the corresponding day, the distributions on two days which are more than 1.5 years apart are very similar,
see black curve (day 561) and blue curve (day 1200, identical to Fig. \ref{fig:Pardus_wealth_cum} A).
Second, after a significant perturbation on day 562 (red curve, after voluntary re-distribution of wealth from the rich to the poor as `Christmas charity'),
the distribution quickly returns to its previous form (green curve: one month after the redistribution).
Comparing the wealth distribution on various days by the Kolmogorov-Smirnov statistic and the Jensen-Shannon divergence,
we find a relaxation time of about 16 days, see Fig. S2 in the SI.

For the timeseries of $g$ and $\alpha$ we find clear anti-correlation, with a Pearson correlation of $\rho=-0.49$
 ($p$-value $<10^{-6}$, ignoring the transient phase in the first 200 days and $2\tau_g$ after the re-distribution). 
The tail of the distribution is neither affected by the charity re-distribution event nor by wars.
An inverse relation for the Gini coefficient and the power-law exponent has also been observed for income in the USA \cite{Banerjee2010}, and is expected to a certain extent. 
The data from Tab. 1 \cite{Banerjee2010} yield $\rho = -0.82$ with a $p$-value $0.001$.
Decreasing $\alpha$ means a more pronounced tail in the wealth distribution, i.e. more extremely rich individuals, resulting in higher inequality, and therefore a higher $g$.

\subsection*{Individual behavioral factors for wealth} \label{sec:behavior}

\subsubsection*{Influence of total activity on wealth} \label{sec:activity}

\begin{figure}[!htbp]
\begin{center}
 \includegraphics[width=0.5\textwidth]{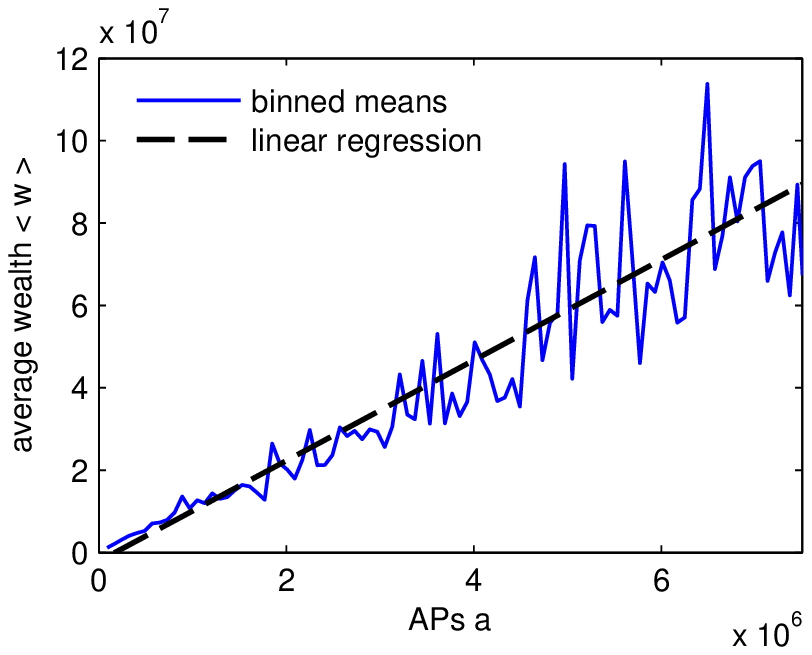}
\end{center}
 \caption{{\bf Wealth as a function of activity $a$.} %
 The blue line shows binned averages of wealth. The dashed line is a linear regression, %
 with a slope $12.2$ credits/AP and a corresponding Pearson correlation coefficient $\rho_{w,a}=0.535$. 
 Data are taken for all active players on day 1200.}
 \label{fig:wealthperAP}
\end{figure}

\begin{figure}[!htbp]
\begin{center}
\includegraphics[width=0.5\textwidth]{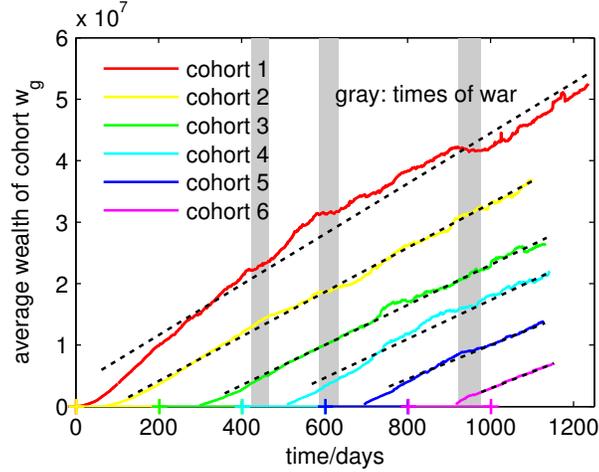}
\end{center}
\caption{{\bf Cohort wealth as a function of time.} Cohort 1 ($G_1$) contains all players who joined Pardus on the first day. 
 Cohort 2 ($G_2$) contains all players you joined between day 2 and 200, cohort 3 ($G_3$) all who joined between day 201 and 400, etc. 
 Wealth $w_{g,j}(t)$ of cohort $G_j$ at time $t$ is calculated as $w_{g,j}(t) = \langle w_i \left(t-t_{0,i}+\bar{t}_{0,j}\right)\rangle_{i \in G_j}$,
 where $t_{0,i}$ is the date at which player $i$ joined the game and $\bar{t}_{0,j} \equiv \langle t_{0,i} \rangle_{i \in G_j}$ is the average cohort entry time. 
 Gray areas mark times of war, dashed lines represent linear fits omitting the transient first 60 days.
 }
 \label{fig:wealthTS}
\end{figure}

We find a trivial strong linear relation between the average wealth of a player and her total activity,  $\langle w \rangle = a \cdot 12.2 -2.09 \times 10^6$, see Fig. \ref{fig:wealthperAP}.
The corresponding Pearson correlation coefficient is $\rho=0.535$ ($p$-value $<10^{-6}$). 
Figure \ref{fig:wealthTS} shows the wealth timeseries of six cohorts of players that joined Pardus during six different time periods.
Cohort 1 contains all players who joined on the first day, cohort 2 joined between day 2 and day 200, cohort 3 between day 201 and 400, etc. 
For each cohort we computed its average wealth from the individual wealth timeseries of its members. 
For the sample, all players present on day 1238 were used. 
Following a short initial phase, average wealth increases almost linearly.
Linear wealth-increase means that players do on average not get better at gaining wealth, i.e. they do not learn over time how to increase their wealth faster.
It is also not consistent with wealth increments proportional to wealth as assumed by the Gibrat model, 
which would instead lead to an exponential wealth-increase on average.
The slopes (i.e. wealth-increase rates) for the different cohorts are different.
We find these slopes to be  4.1, 3.6, 3.3, 3.2, 2.8, and 2.7 $\times10^4$  for cohort 1 to cohort 6, respectively. 
We used a linear fit omitting the first 60 days of each timeseries. 
This means that the older the cohorts, the faster is their average wealth-gain.  
There are two possible interpretations of this result. Either only the players that are more efficient 
in accumulating wealth have stayed in the game to become the old cohorts, or older players occupy the most profitable trades, locations in the game, and younger players 
have no chance to enter these profitable occupied market positions.
We have checked the first interpretation by including all players up to the end of their lifetime,
irrespective of whether or not they were present on day 1238, and found only a marginal effect, see Fig. S3 in the SI.
Effects of war on wealth can be seen.  The wealth of various cohorts stagnates during war and sometimes continues to 
grow with a slightly different slope than before. 
For the younger cohorts, this effect is washed out due to their broad range of entry dates.

\subsubsection*{Wealth and the actions of players}\label{sec:goodBad}

\begin{figure}[!htbp]
\begin{center}
 \includegraphics[width=0.7\textwidth]{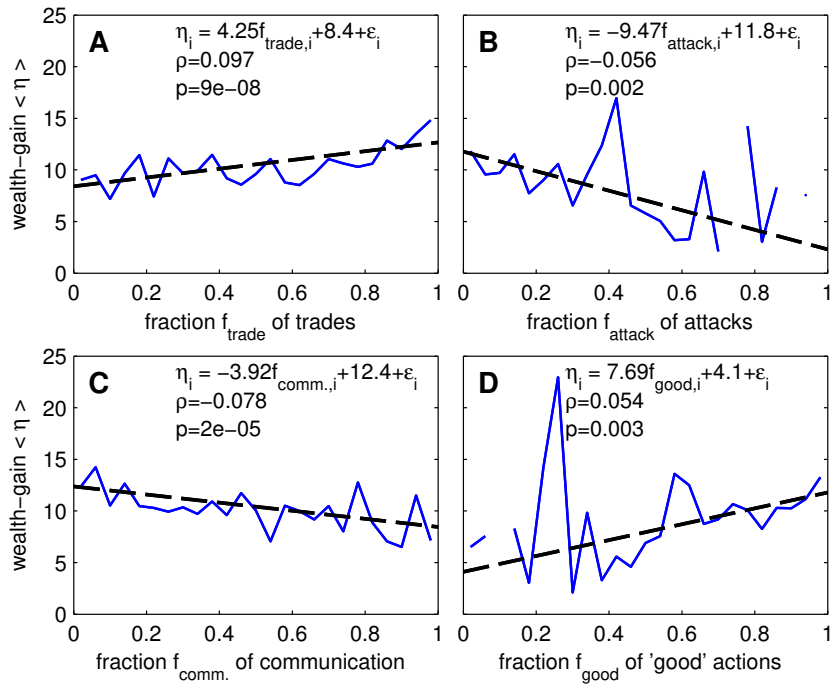}
\end{center}
 \caption{{\bf Wealth-gain as a function of behavior.}
 Behavior is quantified as the fraction of particular actions, such as A trades, B attacks, and C communication,
with respect to all actions performed by an individual.
 The blue line shows binned averages, the dashed line is the linear regression.
 Data consist of all actions between day 1170 and 1200 and wealth-gain on day 1200.}
 \label{fig:behavior}
\end{figure}

Players can interact with each other through trading with each others selling points, communicating, 
directly transferring goods (making gifts\footnote{Called `trade' in \cite{Szell2013,Thurner2012} and merged with trade in \cite{Szell2010msd}}),
 attacking, placing bounties, marking each other as friend or enemy, or removing one of these marks.
Trading, communicating, making gifts, friendship marking, or removing an enemy mark are seen as `cooperative' or `good' actions, while the remaining interactions are destructive or `bad'.
For every player $i$ who is active on day 1200, we count the actions he did since day 1170:
the number of trades he initiated $n_{\mathrm{trade}, i}$,
the number of messages he sent $n_{\mathrm{comm.}, i}$,
the number of gifts he made $n_{\mathrm{gift}, i}$,
the number of attacks he did $n_{\mathrm{attack}, i}$,
the number of bounties he placed $n_{\mathrm{bounty}, i}$,
the number of other players he marks as friend $n_{\mathrm{+friend}, i}$ or enemy $n_{\mathrm{+enemy}, i}$,
the number of friend or enemy marks he removes, $n_{\mathrm{-friend}, i}$ or $n_{\mathrm{-enemy}, i}$.
The total number of activities follows as
 $n_{\mathrm{tot}, i} = n_{\mathrm{trade}, i} + n_{\mathrm{comm.}, i} + n_{\mathrm{gift}, i} + n_{\mathrm{attack}, i} +
 n_{\mathrm{bounty}, i} + n_{\mathrm{+friend}, i} + n_{\mathrm{+enemy}, i} + n_{\mathrm{-friend}, i} + n_{\mathrm{-enemy}, i}$ .
We only consider players with $n_{\mathrm{tot}, i}>0$.
For those we define the fraction of one type of action as
\begin{equation}
 f_{\cdot, i} = \frac{n_{\cdot, i}}{n_{\mathrm{tot}, i}} \quad .
\end{equation}
Accordingly, we define the fraction of `good' actions as
\begin{equation}
 f_{\mathrm{good}, i} = \frac{ n_{\mathrm{trade}, i} + n_{\mathrm{comm.}, i} + n_{\mathrm{gift}, i} + n_{\mathrm{+friend}, i} + n_{\mathrm{-enemy}, i}}{n_{\mathrm{tot}, i}} \quad .
\end{equation}
%
Figure \ref{fig:behavior} A clearly shows that the more a player trades compared to his other actions, the higher is his wealth-gain.
This is no surprise, as trade is the main source of income in the game.
The average fraction of trade is 75.3\%.
Table \ref{tab:PartialCorr} also shows a strong positive partial correlation (see Methods) between trade fraction and wealth.
Figure \ref{fig:behavior} B shows that the more of a player's actions are attacks, the lower is his wealth-gain.
This suggests that revenue from attacks through robbery and bounty hunting does hardly or not exceed the costs for repairing damage done by the fight.
There might be secondary damaging effects of aggressive behavior, such as reduced willingness of others for trade.
A third explanation might be attacks that are carried out without the intent to rob or to collect a bounty, but just for terror.
The average fraction of attacks is 1.7\%.
Table \ref{tab:PartialCorr} shows a significant negative partial correlation between attacks and wealth.
It can be seen in Fig. \ref{fig:behavior} C that players who communicate much have lower wealth-gain.
The main reason for this might be that if a high fraction of the actions consists of sending messages, only a low fraction of actions consists of trades:
while trades are directly influencing wealth, communication is  neutral.
Of course, the same also applies to attacks.
The average fraction of communication is 19.7\%.
Table \ref{tab:PartialCorr} shows a significant negative partial correlation between communication and wealth.
Figure \ref{fig:behavior} D shows that a higher fraction of `good' actions is connected to higher wealth-gain.
The average fraction of `good' actions is 97.5\%.
`Good' actions are mainly trades (which are on average 3/4 of \emph{all} actions and therefore an even higher fraction of \emph{`good'} actions),
while `bad' actions are mainly attacks (on average, 2.5\% of all actions are `bad' while 1.7\% of all actions are attacks).
Therefore, high $f_{\mathrm{good}, i}$ means high $f_{\mathrm{trade}, i}$ and low $f_{\mathrm{attack}, i}$,
both of which are connected to high wealth-gain.
The connection between high fraction of `good' actions and high wealth-gain is a direct consequence.
The partial correlation in Tab. \ref{tab:PartialCorr} also clearly shows a positive partial correlation between the fraction of `good' actions and wealth.

\subsubsection*{Influence of achievement-factors on wealth}

\begin{table*}[p]
\caption{\bf{Partial correlation coefficients for wealth controlling for total activity.}}
 \begin{tabular}{r|*{5}{l}}
 & day 240 & day 480 & day 720 & day 960 & day 1200 \\
\hline
age & $-0.024$ & $-0.043^{**}$ & $-0.037^{**}$ & $-0.040^{**}$ & $-0.064^{****}$ \\
faction rank & $0.157^{****}$ & $0.104^{****}$ & $0.112^{****}$ & $0.107^{****}$ & $0.085^{****}$ \\
XP & $0.189^{****}$ & $0.077^{****}$ & $0.002$ & $-0.036^{*}$ & $-0.056^{****}$ \\
combat skill & $0.063^{****}$ & $0.004$ & $-0.011$ & $-0.036^{*}$ & $-0.088^{****}$ \\
farming skill & $0.010$ & $0.050^{***}$ & $0.086^{****}$ & $0.096^{****}$ & $0.121^{****}$ \\
\hline
$f_\mathrm{trade}$ & $0.008$ & $0.072^{****}$ & $0.058^{***}$ & $0.132^{****}$ & $0.113^{****}$ \\
$f_\mathrm{comm.}$ & $0.011$ & $-0.052^{**}$ & $-0.039^{*}$ & $-0.112^{****}$ & $-0.098^{****}$ \\
$f_\mathrm{attack}$ & $-0.052^{**}$ & $-0.063^{****}$ & $-0.051^{**}$ & $-0.079^{****}$ & $-0.046^{**}$ \\
$f_\mathrm{good}$ & $0.042^*$ & $0.062^{***}$ & $0.058^{***}$ & $0.071^{****}$ & $0.046^{*}$ \\
\hline
$k_\mathrm{in}^\mathrm{trade}$ & $0.245^{****}$ & $0.288^{****}$ & $0.307^{****}$ & $0.300^{****}$ & $0.301^{****}$ \\
$k_\mathrm{out}^\mathrm{trade}$ & $-0.026$ & $0.025$ & $0.037^{**}$ & $0.069^{****}$ & $0.122^{****}$ \\
$C^\mathrm{trade}$ & $-0.050^{***}$ & $-0.052^{***}$ & $-0.058^{****}$ & $-0.050^{***}$ & $-0.052^{***}$ \\
$k_\mathrm{nn}^\mathrm{trade}$ & $-0.074^{****}$ & $-0.069^{****}$ & $-0.056^{****}$ & $-0.062^{****}$ & $-0.056^{****}$ \\
$k_\mathrm{in}^\mathrm{comm.}$ & $0.054^{***}$ & $0.035^{*}$ & $0.036^{**}$ & $0.036^{*}$ & $0.099^{****}$ \\
$k_\mathrm{out}^\mathrm{comm.}$ & $0.033^{*}$ & $0.026$ & $0.034^{*}$ & $0.045^{**}$ & $0.095^{****}$ \\
$C^\mathrm{comm.}$ & $-0.022$ & $-0.012$ & $-0.010$ & $-0.012$ & $-0.000$ \\
$k_\mathrm{nn}^\mathrm{comm.}$ & $-0.032^{*}$ & $-0.053^{****}$ & $-0.028^{*}$ & $-0.041^{**}$ & $-0.024$ \\
$k_\mathrm{in}^\mathrm{friend}$ & $0.054^{***}$ & $0.023$ & $-0.027$ & $-0.016$ & $-0.016$ \\
$k_\mathrm{out}^\mathrm{friend}$ & $0.022$ & $-0.009$ & $-0.055^{****}$ & $-0.053^{***}$ & $-0.069^{****}$ \\
$C^\mathrm{friend}$ & $0.030^{*}$ & $0.010$ & $0.008$ & $0.012$ & $0.010$ \\
$k_\mathrm{nn}^\mathrm{friend}$ & $-0.040^{**}$ & $-0.035^{*}$ & $-0.011$ & $-0.007$ & $-0.015$ \\
$k_\mathrm{in}^\mathrm{enemy}$ & $-0.079^{****}$ & $-0.077^{****}$ & $-0.083^{****}$ & $-0.087^{****}$ & $-0.029^{*}$ \\
$k_\mathrm{out}^\mathrm{enemy}$ & $0.040^{**}$ & $0.023$ & $0.036^{**}$ & $0.035^{*}$ & $0.025$ \\
$C^\mathrm{enemy}$ & $0.017$ & $-0.018$ & $-0.007$ & $-0.017$ & $0.026$ \\
$k_\mathrm{nn}^\mathrm{enemy}$ & $0.058^{****}$ & $0.023$ & $0.035^{*}$ & $0.051^{***}$ & $0.029^{*}$ \\
 \end{tabular} 
\begin{flushleft}
 Data taken at days 240, 480, 720, 960, and 1200 after the beginning of the game. %
${}^*\;p-\mathrm{value}<0.05, \;{}^{**} \;p-\mathrm{value}<0.01, \;{}^{***} \;p-\mathrm{value}<0.001, \;{}^{****} \;p-\mathrm{value}<0.0001$.
\end{flushleft}
\label{tab:PartialCorr}
\end{table*}

\begin{figure}[!htbp]
\begin{center}
 \includegraphics[width=\textwidth]{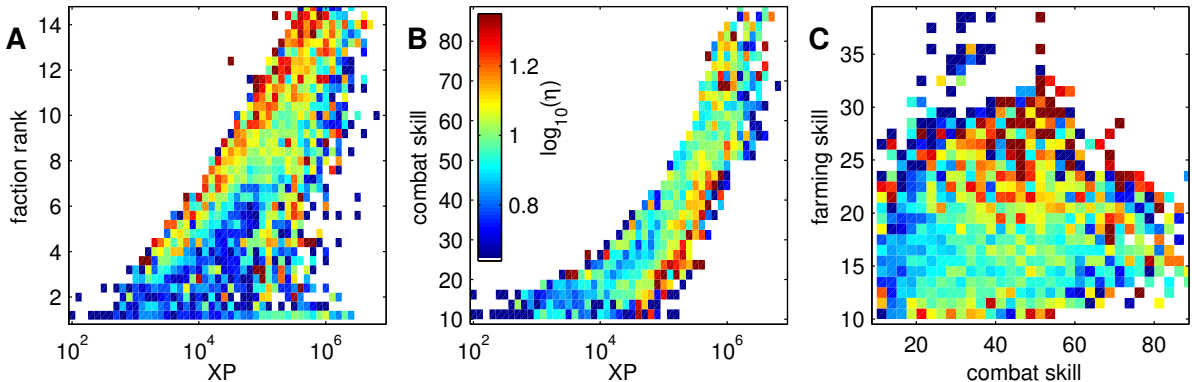}
\end{center}
\caption{{\bf 2D binned averages of the wealth-gain as a function of achievement-factors.} 
Colors represent the logarithm of the average wealth-gain, $\log_{10}(\eta)$, over all players that fall into that bin.
Blue corresponds to low, red to high values, empty bins are white.
A XP and faction rank, B XP and combat skill, C combat skill and farming skill.
Data are taken every 240 days (see Methods).}
 \label{fig:OtherProperties2D_w}
\end{figure}

Wealth as well as the achievement-factors, such as skills, XPs, and faction rank, are strongly correlated with total activity.
To exclude these spurious correlations, partial correlation coefficients are calculated (see Methods) and reported in the upper part of Tab. \ref{tab:PartialCorr}.
We take snapshots of the achievement-factors and of the wealth on days 240, 480, 720, 960, and 1200 respectively.
Stars in Tab. \ref{tab:PartialCorr} indicate the significance level for the null-hypothesis that the given coefficient is zero.
A correlation model over all variables (see Tab. S2 in SI) does not yield improvement over the one-dimensional regression of $w$ on $a$ as shown in Fig. \ref{fig:wealthperAP}.
To examine possible nonlinear relations between achievement-factors and wealth, in Fig. \ref{fig:OtherProperties2D_w} we show two-dimensional binned averages of wealth-gain
as a function of faction rank, XP, combat- and farming skill.
To produce two-dimensional binned averages, players were sorted into bins according to two achievement-factors.
For every bin the average wealth-gain $\eta$ (see Eq. \ref{eq:defEta}) of all players in that bin is determined and represented as the color of the bin.
If no player with a certain combination of achievement-factors is found, the corresponding bin is empty, and the bin color is white.

From Fig. \ref{fig:OtherProperties2D_w} and Tab. \ref{tab:PartialCorr} we find the influence and significance of the various factors:
\begin{description}
 \item[Age] is a significant factor (significance level below 1\%) at four out of the five time points.
The negative coefficient seems to be in contrast to the increase of wealth with age seen in Fig. \ref{fig:wealthTS}.
The explanation is that total activity, which is most strongly correlated with wealth, is limited by age.
This induces the spurious correlation between wealth and age seen in Fig. \ref{fig:wealthTS}.
 \item[Faction rank] is a significant positive factor for wealth with a significance level below 0.01\% for all days. High faction rank means `political' influence in the game.
Players that are in no faction, i.e. less social, have the smallest possible value as faction rank and are on average poorer.
Figure \ref{fig:OtherProperties2D_w} A shows that a high faction rank correlates strongly with wealth-gain. 
We also see that the non-empty bins suggest a strong correlation between XP and faction rank. 
 \item[XP] is significantly positive for wealth on the first two sample days with continually decreasing coefficient, changing sign on the last two days.
This might indicate that XP is positive up to a certain extent, after which the goal of high XP starts to be in contradiction to the goal of high wealth.
In Fig. \ref{fig:OtherProperties2D_w} A, the data from all five days are combined, and the positive and negative correlations of XP cancel and leave no significant effect of XP on wealth-gain.
 \item[Combat skill] has a correlation with wealth similar to XP, see Tab. \ref{tab:PartialCorr}.  
We see in Fig. \ref{fig:OtherProperties2D_w} B that combat skill is approximately proportional to the logarithm of XP.
There is a significant fraction of rich people with low combat skill of about 20.
Figure \ref{fig:OtherProperties2D_w} C shows no correlation between combat skill and wealth-gain.
 \item[Farming skill] has a consistently positive and mostly significant correlation with wealth. Farming skill is associated with the collection of resources, which generates income.
Figure \ref{fig:OtherProperties2D_w} C also suggests an association between high farming skill and high wealth-gain.
\end{description} 

\subsection*{The effects of groups on wealth: the value of being social}

\begin{figure}[!htbp]
\begin{center}
  \includegraphics[width=\textwidth]{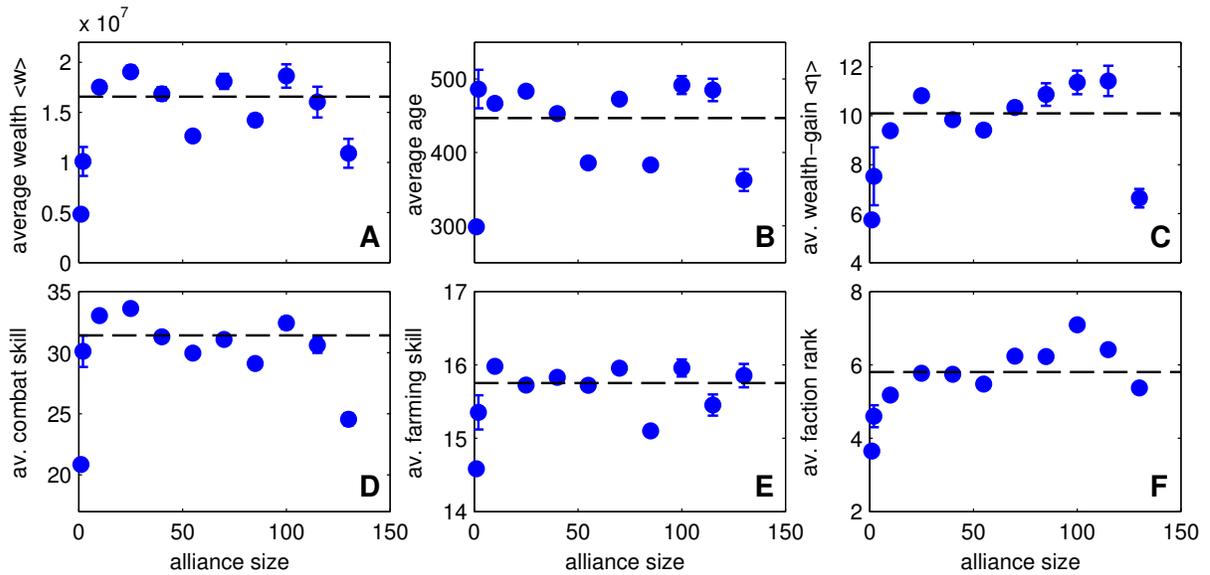}
\end{center}
 \caption{{\bf Wealth and other properties as a function of alliance size.} 
 We bin players according to the size of the alliance they belong to, and show various properties as a function of alliance size:
A wealth, B age, C wealth-gain, D combat skill, E farming skill, F faction rank.
First bin are players in no alliance, second bin are players in an alliance of size two.
Clearly members of these smallest alliances show low wealth and achievement measures.
Also for the largest groups, lower levels are observable.
Error bars denote the standard errors of these means (assuming Gaussian distributions). 
The black dashed line shows the average over all players in an alliance with at least three members. 
Data are taken every 240 days (see Methods).}
 \label{fig:alliSize}
\end{figure}

\begin{table}[!ht]
\caption{\bf{Various properties of players depending on whether they are member in an alliance or not.}}
\begin{tabular}{l|cc}
 Average & no alliance & alliance \\
\hline
wealth $w$ & $4.83\times 10^{6}$ & $1.66\times 10^{7}$\\
age & $299$ & $447$\\
total activity $a$ & $6.53\times 10^{5}$ & $1.61\times 10^{6}$ \\
wealth-gain $\eta$ & $5.75$ & $10.09$ \\
combat skill & $20.9$ & $31.4$ \\
farming skill & $14.6$ & $15.8$ \\
faction rank & $3.65$ & $5.81$ \\
\end{tabular}
 \begin{flushleft}
All $p$-values obtained from a two sample $t$-test and a Wilcoxon rank sum test are less than $10^{-6}$. 
Data are taken every 240 days (see Methods).
\end{flushleft}
\label{tab:alliNoall}
\end{table} 

Players in Pardus organize within social groups that are called alliances in the game.
At day 1200, 161 alliances with an average size of 23 members exist. Being a member of an alliance is a social 
commitment. 
In Tab. \ref{tab:alliNoall} we collect the average values for several features of players, 
depending on whether they are alliance members or not.
In general, alliance members are richer, both in absolute terms and in terms of wealth-gain than those that are not alliance members.
Members also have better skills and a higher faction rank. 
In Fig. \ref{fig:alliSize} we see that the size of an alliance has little influence on wealth and other factors,
 except for players that are in alliances with only two members.
These are consistently poorer than the players in groups with three or more members.
Members of the biggest alliances also have some indicators below the average (dashed line).

\subsection*{The effects of social networks on wealth}

\begin{figure}[!htbp]
\begin{center}
 \includegraphics[width=\textwidth]{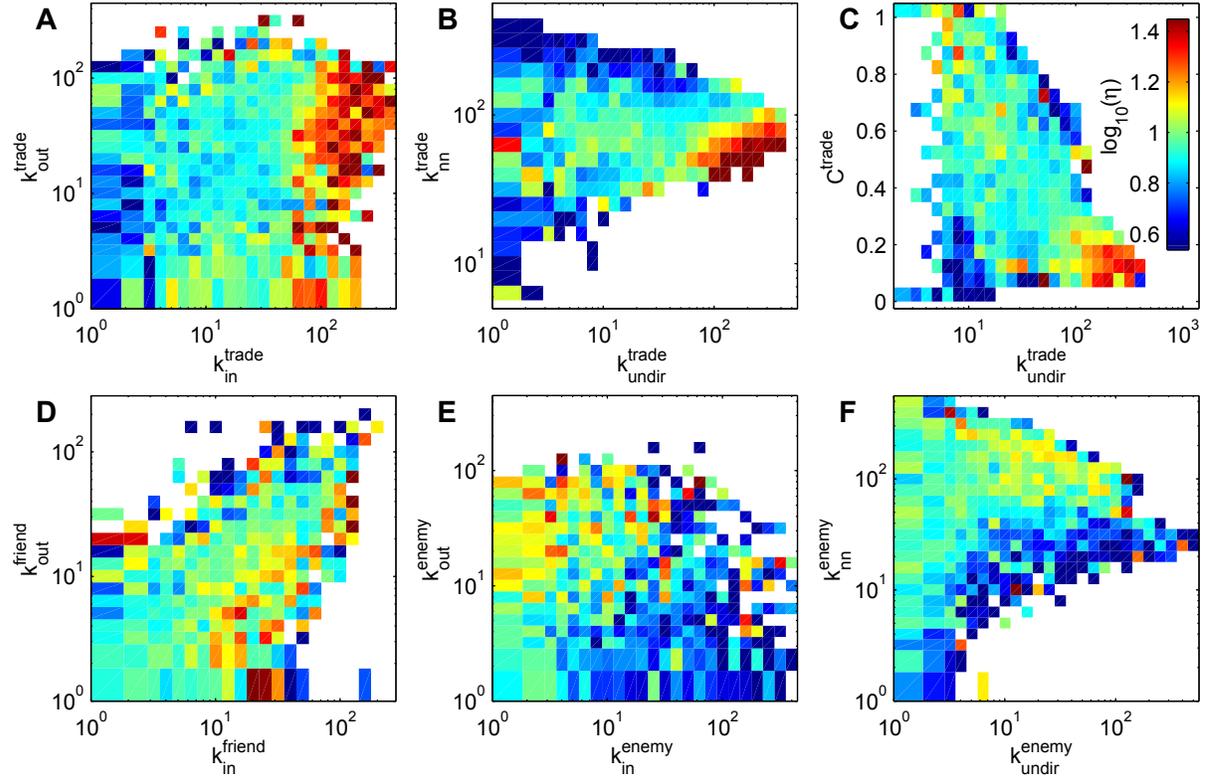} 
\end{center}
 \caption{{\bf Wealth-gain as a function of network properties.} %
Color represents the logarithm of the wealth-gain, $\log_{10}(\eta)$, from blue (lowest) to red (highest), empty bins are white. 
A trade in- and out-degree, B trade undirected degree and nearest-neighbor degree, C trade undirected degree and clustering coefficient,
D friend in- and out-degree, E enmity in- and out-degree, F enmity undirected degree and nearest-neighbor degree.
Data are taken every 240 days (see Methods).}
 \label{fig:NWProperties2D_w}
\end{figure}

We use the trade, communication, friendship, and enemy networks of Pardus (see SI), which are 
available for every day. For every node (player) $i$ we determine the in- and out-degree ($k_{\mathrm{in},i}, k_{\mathrm{out},i}$), 
the nearest-neighbor degree $k_{\mathrm{nn},i}$, and its clustering coefficient (see SI).
We calculate partial correlations between wealth and the network parameters controlling for total activity.
We collect the results in the lower part of Tab. \ref{tab:PartialCorr}. 
To elucidate the dependence of wealth on various combinations of network factors, 
in Fig. \ref{fig:NWProperties2D_w} we plot two-dimensional binned averages of wealth-gain versus pairs of network properties.
The results are:

\begin{description}
 \item[Trade network] As expected, the trade network has the strongest influence on wealth. 
Trade in-degree has a significant, positive partial correlation with wealth.
The in-degree is defined as trade with a player's production facilities and is therefore a proxy for his production.
Figure \ref{fig:NWProperties2D_w} A confirms the positive connection between trade in-degree and wealth, while not showing any influence of trade out-degree.
 However, Tab. \ref{tab:PartialCorr} reports a positive correlation between wealth and active trade with the production facilities of fellow players,
in agreement with the positive effect of active trading shown in Fig. \ref{fig:behavior} A.
Figure \ref{fig:NWProperties2D_w} B presents the undirected degree of the trade network versus the nearest-neighbor degree.
The richest are found to have an intermediate nearest-neighbor degree of about $k^\mathrm{trade}_\mathrm{nn} \sim 35-70$, well below their undirected degree.   
This means that they are selling to people that are less connected in the trade network than they are themselves. 
Table \ref{tab:PartialCorr} shows a negative correlation between the nearest-neighbor degree and wealth with a significance level below 0.01\%.
From Fig. \ref{fig:NWProperties2D_w} C we gather that high wealth-gain is made with a combination of high degree and 
a relatively low clustering coefficient, $C^\mathrm{trade} \sim 0.1$.  This means that rich players avoid cyclical structures in their trading 
networks, which allows them to act as `brokers' between players that do not directly trade with each other.
The partial correlation coefficient between wealth and the trade clustering coefficient is negative.

 \item[Communication network] Communication in-degree has a significantly positive partial correlation coefficient.
High communication in-degree means good access to information, which is expected to be profitable.
The Communication out-degree shows positive partial correlation on most days.
A player's communication out-degree is the number of fellow players she \emph{tries} to influence.
Since most communication links are reciprocal, and in- and out-degree are therefore highly correlated, there might be a spurious effect of the communication in-degree.
 The communication nearest-neighbor degree has a negative and mostly significant partial correlation coefficient.
This might indicate it is advantageous to mainly converse with fellow players who are less informed than oneself.

 \item[Friendship network] In Fig. \ref{fig:NWProperties2D_w} D the situation for the in- and out-degrees for the friendship network is shown. 
It is visible that players with high wealth-gain are those that are liked by more players than they like themselves, 
$k_\mathrm{in}^\mathrm{friend} > k_\mathrm{out}^\mathrm{friend}$.  Poor players have marked fellow players as friends more often on average 
than they have been marked.
 In Tab. \ref{tab:PartialCorr}, friendship in-degree hardly shows any correlation with wealth,
 while friendship out-degree has a significant negative correlation on all sampling days except day 240.
This might indicate that time and resources invested into friendship are missing for the generation of wealth.

 \item[Enmity network] We see that people with above average wealth-gain are very rarely marked as an enemy by others,
 but do mark others as enemies, see Fig. \ref{fig:NWProperties2D_w} E. 
Players who have been marked as enemy by many others are generally poor.
In agreement with this finding, the enmity in-degree has a significant negative partial correlation with wealth, 
while the enmity out-degree has a weak significant positive correlation with wealth, see Tab. \ref{tab:PartialCorr}.
This suggests that players with high wealth-gain actively invest in a good reputation. 
Finally, players with above average wealth-gain have a high nearest-neighbor degree, see Fig. \ref{fig:NWProperties2D_w} F.
Table \ref{tab:PartialCorr} reports (mainly) significant positive correlations between wealth and the enmity nearest-neighbor degree.
Players with high enmity (in)degree are `public enemies' \cite{Szell2010msd}.
A high $k_\mathrm{nn}^\mathrm{enemy}$ means that one is mainly the enemy of public enemies and that one has few private enemies.
\end{description}

\section*{Discussion}

We studied the economy of the virtual world of the MMOG \emph{Pardus}.
We found that the wealth distribution in Pardus has a similar shape like wealth distributions of `real' countries,
including an exponential bulk and a power-law tail.
The power-law exponent of Pardus is within the range of real-world power-law exponents describing the moderately rich.
The Gini index shows that wealth is slightly more equally distributed in Pardus than in many Western industrial countries.
We observed that the shape of the wealth distribution is stable: eventual external perturbations exponentially relax to the stable state.
While the total wealth in the Pardus game increases over time, large scale conflicts hamper the creation of wealth.
We found that an average player's wealth grows linearly with his total activity.
As total activity is limited by a player's age (time in the game), wealth also increases linearly with the age of a player.
Linear increase suggests that neither learning nor proportional growth (i.e. `rich get richer') are dominant on a global scale.
Players who entered the game earlier have higher wealth-increase rates.

For the first time, we could observe the connections between personal wealth and social behavior.
We found that wealthy players organize in social groups.
A group size between three and 120 members appeared to be best for wealth and achievement-factors.
We found that wealthy players invest in their social reputation by constructive actions.
Personal wealth in Pardus is connected to skills for collecting resources and high `political' influence,
but not to combat skill and fighting experience.
Analyzing the trade network, we observed that wealthy players trade with many others, while their trade partners trade with fewer others, and hardly among each other.
Taken to the extreme, the wealthy organize their local trade network so that they are the hub of a star-like network.
In the friendship and enmity networks we observed that the wealthy are well respected,
 and show animosity -- if at all -- only towards public enemies.

\section*{Materials and Methods}
\subsection*{Datasets}\label{sec:data}
We study data from one of three game universes of Pardus, \emph{Artemis}.
Days are counted from the opening of this server, day one is June 12, 2007.
For dataset 1, used for Figs. \ref{fig:Pardus_wealth_cum}, \ref{fig:wealthperAP}, and \ref{fig:behavior}, 
we extract snapshot data on day 1200 since the opening of the game, i.e. September 23, 2010.
We selected only those players that have been active in the last 30 days before day 1200.
For dataset 2, used for Fig. \ref{fig:ginietc_evolution}, we extracted data on every day and applied the same filtering as for dataset 1,
 i.e. we excluded players whose last activity was longer than 30 days ago.
For dataset 3, used for Fig. \ref{fig:wealthTS}, we took the complete timeseries of all players that were in the game on day 1238,
which is the last day included in our database.
For dataset 4, used for Figs. \ref{fig:NWProperties2D_w} and \ref{fig:alliSize} and Tabs. \ref{tab:PartialCorr} and \ref{tab:alliNoall},
we used snapshot datasets separated by 240 days, starting at day 240.
After 240 days, the autocorrelation function of wealth has decayed to $\rho^\mathrm{auto}=0.355$, so the single data points can be treated as independent.
The data contain a daily snapshot of the friendship and enmity networks, all players' possessions,
and alliance membership.
For the trade network, we draw a link on day $t^*$ if a trade has taken place in the time range $[t^*-60, t^*]$.
Players who have only recently joined the game are naturally close to their initial wealth
and are therefore excluded from datasets 1, 2, and 4.
As a criterion for admitting a player to the dataset, we require that the players have actively played for ten days,
more precisely: they have spent at least 50,000 APs.
Dataset 1 contains 3,245 players, dataset 2 contains 4,483,175 data points from 16,662 distinct players, dataset 3 contains 3,693 players,
and dataset 4 contains 25,195 data points from 12,186 distinct players on 5 distinct days.
Dataset 1 is a proper subset of dataset 2 and also of dataset 4.


\subsection*{Lorenz curve and Gini index}\label{sec:defGini}
Let $N$ be the number of players, and $w_i$ the wealth of player $i$, ordered so that $w_i \leq w_{i+1} \; \forall \; i \in \{1 \ldots N \}$.
The Lorenz curve consists of points 
\begin{equation*}
L^x_j=\frac j N \quad, \qquad L^y_j=\frac{\sum\limits_{i=1}^j w_i}{\sum\limits_{i=1}^N w_i} \quad , 
\end{equation*}
and their piecewise linear connection.
For complete equality, $w_i=w_j\forall \; i,j \in \{1 \ldots N \}$, $w_i$ cancels and $L^x_j=L^y_j$, turning the Lorenz curve into a straight line from $(0,0)$ to $(1,1)$.

Let $A$ be the area under the Lorenz curve. The Gini index \cite{Gini1912} is defined as $g \equiv 1-2A$.
It can be calculated from the data by:
\begin{equation*}
 g=1-2\left(\frac{\sum\limits_{i=1}^N \; (N+1-i)w_i}{N\sum\limits_{i=1}^N w_i} -\frac 1 {2N} \right) \quad .
\end{equation*}
For complete equality, $g=0$, and for maximal inequality ($w_i=0$ for $i<N$), $g=1-\textstyle\frac 1 N$. 

\subsection*{Correlation coefficients and partial correlations}\label{sec:defCorrelation}
Throughout the paper we report correlations by the widely used Pearson's correlation coefficient, calculated from data as \cite{Hartung1999}:
\begin{equation*}
 \rho_{w,x} = \frac{\sum\limits^N_{i=1}(w_i - \langle w \rangle)(x_i - \langle x \rangle)}{\sqrt{\sum \limits^N _{i=1}(w_i - \langle w \rangle)^2 \sum\limits^N _{i=1}(x_i - \langle x \rangle)^2}} \quad ,
\end{equation*}
where $\langle \cdot \rangle$ denotes the average over all $i$.
To determine the effect of single factors on wealth while removing the effect of total activity, we calculate \emph{partial correlations} controlling for total activity:
For both wealth $w$ and the studied factor $x$, the linear regression on total activity $a$ is calculated.
The correlation between residuals of these regressions is the partial correlation coefficient $\rho_{(w,x)/a}$ \cite{Hartung1999}.
Equivalently, $\rho_{(w,x)/a}$ can more easily be calculated as:
\begin{equation*}
 \rho_{(w,x)/a} = \frac{\rho_{wx}-\rho_{wa}\rho_{xa}}{\sqrt{(1-\rho_{wa}^2)(1-\rho_{xa}^2)}} \quad .
\end{equation*}

\section*{Acknowledgments}
The authors acknowledge support from the Austrian Science Fund FWF P23378 and from FP7 project CRISIS.



\end{document}


\begin{flushleft}
{\Large
\textbf{Supporting Information}
}
\end{flushleft}

This document is the Supplementary Information of the manuscript ``Behavioral and Network Origins of Wealth Inequality: Insights from a Virtual World''.
It contains a collection of real-world data for comparison, a definition of network measures, and further information.

\tableofcontents

\section{Power Law Exponents for Tails of Wealth Distributions} \label{sec:RLPareto}
\begin{table}[!ht]
\caption{
\bf{Comparison of power law exponent $\alpha$ of the Pardus wealth distribution to real world data}}
  \begin{tabular}{lccll}
   country & year & $\alpha$ & method & source \\
   Pardus & 2010 & $2.46$ & data from MMOG & own fit \\
   Ancient Egypt & ca. 1380 BC & $3.76 \pm 0.22$ & excavation data & \cite{Abulmagd2004} \\
   Hungary & 1550 & 0.92 & historical almanac & \cite{Hegyi2007} \\
   Hungary & 1767 -- 1773 & 0.99 & historical almanac & \cite{Hegyi2007} \\
   Sweden & 1931 & 1.5 & wealth tax & \cite{Steindl1965} \\
   Sweden & 1959 & 1.7 & wealth tax & \cite{Steindl1965} \\
   India & 1991 & 2.04 -- 2.44 & survey & \cite{Jayadev2008} \\
   India & 2002 & 1.85 -- 2.17 & survey & \cite{Jayadev2008} \\
   France & 1994 & $1.82 \pm 0.03$ & encyclopedia & \cite{Levy1998} \\
   UK & 1996 & 1.9 & inheritance tax & \cite{Dragulescu2001} \\
   UK & 1997 & $1.06 \pm 0.004$ & ranking of wealthiest & \cite{Levy1998} \\
   Sweden & 1999 & $1.54 \pm 0.05$ & wealth tax & own fit, see S\ref{fig:Sweden} \\
   Sweden & 2000 & $1.58 \pm 0.02$ & wealth tax & own fit, see S\ref{fig:Sweden} \\
   Sweden & 2001 & $1.64 \pm 0.02$ & wealth tax & own fit, see S\ref{fig:Sweden} \\
   Sweden & 2002 & $1.61 \pm 0.04$ & wealth tax & own fit, see S\ref{fig:Sweden} \\
   Sweden & 2003 & $1.61 \pm 0.03$ & wealth tax & own fit, see S\ref{fig:Sweden} \\
   Sweden & 2004 & $1.61 \pm 0.04$ & wealth tax & own fit, see S\ref{fig:Sweden} \\
   Sweden & 2005 & $1.62 \pm 0.04$ & wealth tax & own fit, see S\ref{fig:Sweden} \\
   Sweden & 2006 & $1.59 \pm 0.05$ & wealth tax & own fit, see S\ref{fig:Sweden} \\
   Sweden & 2007 & $1.63 \pm 0.04$ & wealth tax & own fit \\
   USA & 1988 -- 2003 & 1.1 -- 1.7 & ranking of wealthiest & \cite{Klass2007} \\
   India & 2002 & 0.81 & ranking of wealthiest & \cite{Sinha2006} \\
   India & 2004 & 0.92 & ranking of wealthiest & \cite{Sinha2006} \\
   China & 2003 & 2.285 & ranking of wealthiest & \cite{Ning2007} \\
   China & 2004 & 2.043 & ranking of wealthiest & \cite{Ning2007} \\
   China & 2005 & 1.758 & ranking of wealthiest & \cite{Ning2007} \\
  \end{tabular} 
\label{tab:Wealth_exponent}
\end{table}

\begin{figure}[!ht]
\begin{center}
\includegraphics[width=17.35cm]{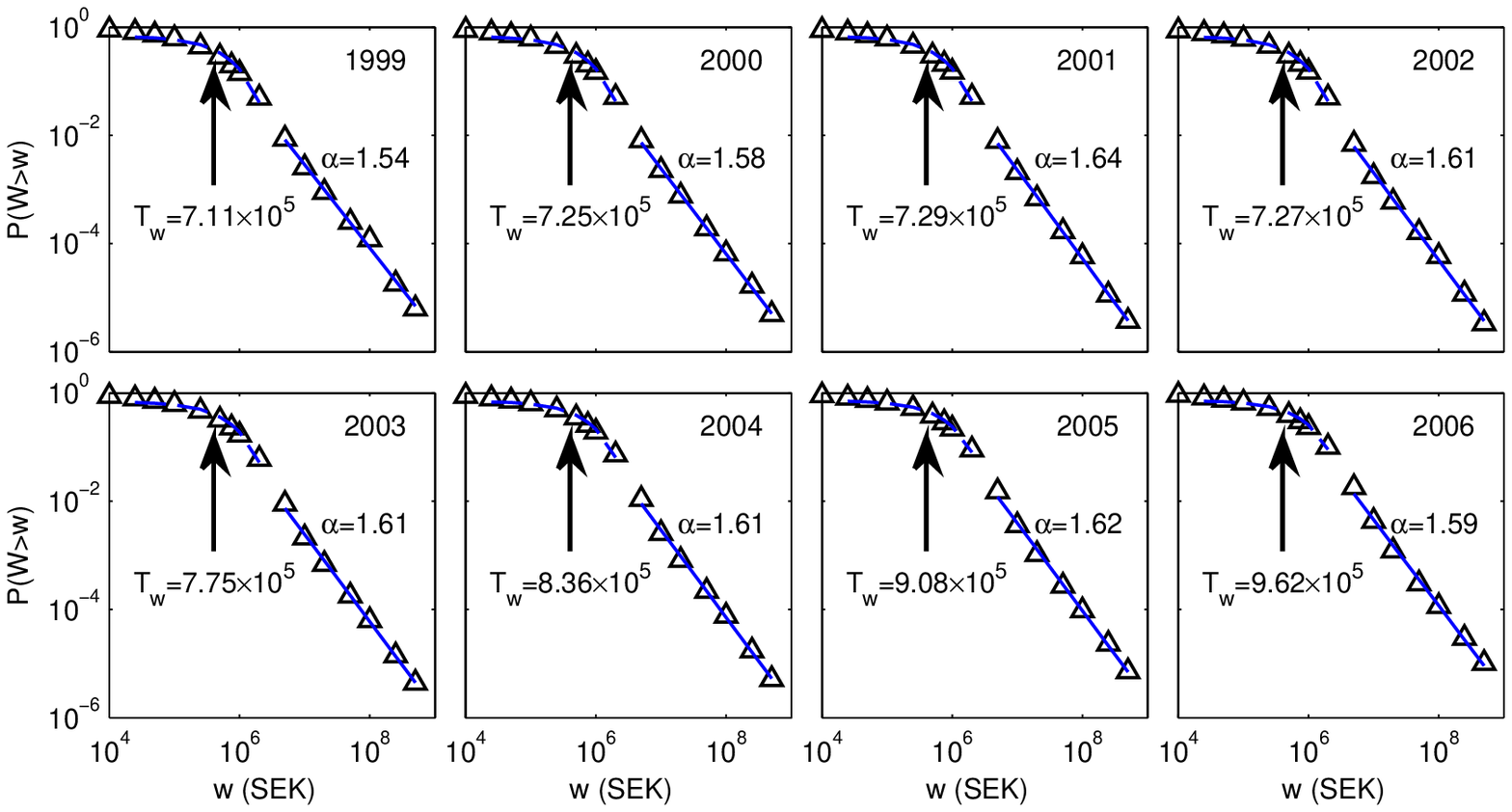}
\end{center}
\caption{
{\bf Wealth distribution of Sweden from 1999 to 2006 with fits.}
Black triangles mark the data, the continuous blue line is a power law fit with exponent as indicated, %
 and the broken blue line is an exponential fit with 'wealth temperature' as indicated.
Data source: Statistics Sweden (2010). Number of women and men with net wealth in different intervals 1999--2007, corrected 2010-03-22.
Available: http://www.scb.se/en\_/Finding-statistics/Statistics-by-subject-area/Household-finances/Income-and-income-distribution/Households-assets-and-debts/Aktuell-Pong/2007A01K/Time-series-tables-19992007/Number-of-women-and-men-with-net-wealth-in-different-intervals-19992007-Corrected-2010-03-22/.
Accessed~11~February~2014.}
\label{fig:Sweden}
\end{figure}

The power law fit in Fig. S\ref{fig:Sweden} is a simple linear least-squares fit to $\log(P(W>w))(log(w))$.

\section{Stability of the Wealth Distribution}
\begin{figure}[!ht]
\begin{center}
\includegraphics[width=8.3cm]{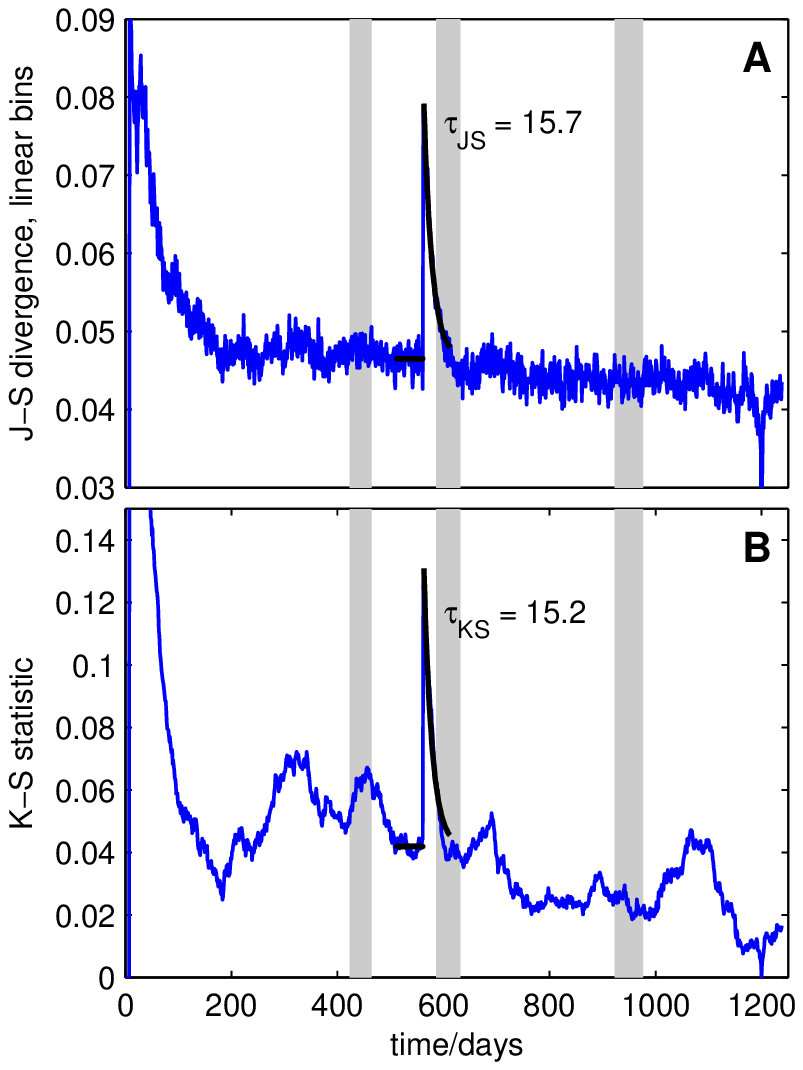}
\end{center}
\caption{
{\bf Comparison between renormalized wealth distribution on day 1200 and on a varying second day.} 
A Jensen-Shannon divergence, B Kolmogorov-Smirnov statistic. 
Black curves fit the decay of the perturbation by an exponential with decay time A $\tau_\mathrm{JS}=15.7$ and B $\tau_\mathrm{KS}=15.2$. Dotted black lines mark the previous level.
}
\label{fig:JS_KS}
\end{figure}

To further quantify the relaxation of the wealth distribution after the perturbation on day 562, we compare the daily wealth distribution to the distribution shown in Fig. 1.
We know from Fig. 2A that wealth grows over time and are interested in the shape of the distribution.
For this reason, we rescale wealth by the daily mean wealth $\langle w(t) \rangle$ like for Fig. 2D.
The rescaled distributions are compared by A the Jensen-Shannon divergence and B the Kolmogorov-Smirnov statistic.
Both measures clearly show a peak on day 562 which decays exponentially with decay time A $\tau_\mathrm{JS}=15.7$days and B $\tau_\mathrm{KS}=15.2$days.

\subsection{Jensen-Shannon Divergence} \label{sec:JS}
The Jensen-Shannon divergence is an information theoretic measure to compare two (probability) distributions.
It is a symmetrized and version of the Kullback-Leibler divergence.
The Jensen-Shannon divergence of two distributions $p$ and $q$ is defined as:
\begin{equation*}
 \mathrm{JSD}(p \parallel q) = \frac{1}{2}D(p \parallel m)+\frac{1}{2}D(q \parallel m)
\end{equation*}
where $m$ is the mean distribution $m=\frac 1 2 (p+q)$ and $D(p \parallel m)$ is the Kullback-Leibler divergence:
\begin{equation*}
D(p\parallel m) = \sum\limits_{i=0}^N \ln\left(\frac{p_i}{m_i}\right) p_i
\end{equation*}
Here, $p_i$ was defined as the probability that the wealth of one randomly selected individual was in interval (bin) $I_i$,
with $I_i$ being intervals of uniform length, $I_0$ having $0$ as lower limit, $N=3,536$, and $I_N$ having an upper limit of $17.68$.

\subsection{Kolmogorov-Smirnov Statistic}
The Kolmogorov-Smirnov statistic compares two cumulative distribution functions $P(x)=\int_x^\infty p$ and $Q(x)=\int_x^\infty q$ by giving their largest probability difference:
\begin{equation*}
D_n=\sup_x |P(x)-Q(x)|
\end{equation*}

\section{Lifetime Bias}

\subsection{Wealth of Cohorts}

\begin{figure}[!ht]
\begin{center}
\includegraphics[width=8.3cm]{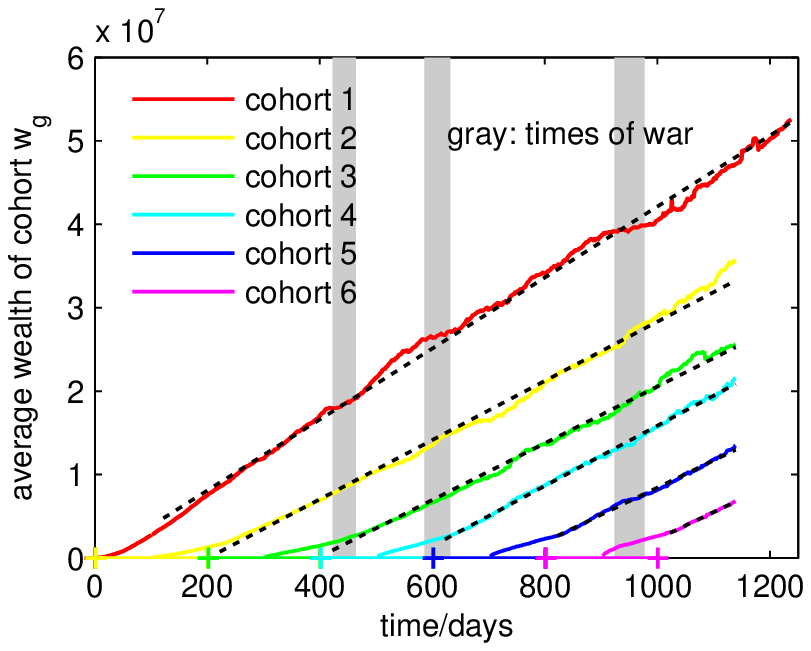}
\end{center}
\caption{{\bf Cohort wealth as a function of time.} Cohort 1 ($G_1$) contains all players who joined Pardus on the first day. 
 Cohort 2 ($G_2$) contains all players you joined between day 2 and 200, cohort 3 ($G_3$) all who joined between day 201 and 400, etc. 
 Wealth $w_{g,j}(t)$ of cohort $G_j$ at time $t$ is calculated as $w_{g,j}(t) = \langle w_i \left(t-t_{0,i}+\tilde{t}_{0,j}\right)\rangle_{i \in G_j(t)}$,
 where $t_{0,i}$ is the date at which player $i$ joined the game and $\tilde{t}_{0,j} \equiv \left( \min_{i \in G_j} \left( t_{0,i} \right) + \max_{i \in G_j} \left( t_{0,i} \right) \right) /2$ is the average cohort entry time. 
 Players are considered as long as they are in the game, i.e. the sizes of the cohorts are not fixed, but may decrease over time,
 $G_j(t) \subseteq G_j(t+1) \; \left( t \geq \max_{i \in G_j} \left( t_{0,i} \right) \right)$.
Gray areas mark times of war, dashed lines represent linear fits omitting the transient first 120 days.
 }
\label{fig:wealthTS}
\end{figure}

Figure S\ref{fig:wealthTS} shows the wealth timeseries of six cohorts of players that joined Pardus during six different time periods.
Cohort 1 contains all players who joined on the first day, cohort 2 joined between day 2 and day 200, cohort 3 between day 201 and 400, etc. 
For each cohort we computed its average wealth from the individual wealth timeseries of its members.
The sample contains all players who have at any time spent 50,000 APs for their entire lifetime.
After a short initial phase (approximately 120 days, as many players leave within 120 days), average wealth increases almost linearly, 
but with different slopes for the different cohorts.
We find these slopes to be  4.3, 3.5, 3.4, 3.6, 3.3, and 3.2 $\times10^4$  for cohort 1 to cohort 6, respectively.

\subsection{Probability to Leave} 

\begin{figure}[!ht]
\begin{center}
\includegraphics[width=8.3cm]{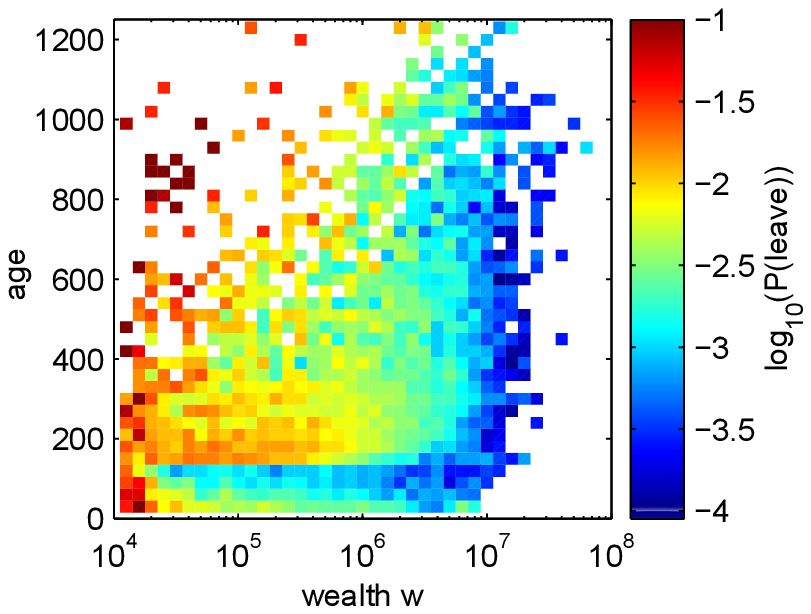}
\end{center}
\caption{{\bf Probability to leave the game as function of age and wealth.} 
Every day except the last, players are sorted into bins according to their current age and their current wealth:
$N_\mathrm{tot}(w,\mathrm{age}) = \sum_t \#\{ i : \log_{10}(w_i(t)) \in \left] \log_{10}(w)-\delta_w, \log_{10}(w)+\delta_w \right] \wedge t_{0,i} \in \left] t-\mathrm{age}-\delta_\mathrm{age}, t-\mathrm{age}+\delta_\mathrm{age} \right] \}$,
where the $\delta$s denote half the bin size (and all other quantities as in the caption of Fig. S\ref{fig:wealthTS}).
In a similar way, we count players per bin that are not in the game anymore the next day, $N_\mathrm{leave}(w,\mathrm{age})$.
The frequency (empirical probability) for a player with a certain combination of wealth and age to leave the game is than $P(\mathrm{leave} | w,\mathrm{age}) = N_\mathrm{leave}(w,\mathrm{age})/N_\mathrm{tot}(w,\mathrm{age})$.
Color in the plot denotes $log_{10}\left(P(\mathrm{leave} | w,\mathrm{age})\right)$, bins with insufficient data, i.e. $N_\mathrm{leave}(w,\mathrm{age})=0$, are colored white.
Only players who had `spent' at least 50,000 APs have been taken into account.}
\label{fig:deathprob}
\end{figure}
Figure S\ref{fig:deathprob} clearly shows that the richer a player is, the lower the probability that he will leave the game.
A second very prominent feature is a step in the probability to leave after day 120.
This is effected by the game mechanics, automatically deleting players after 120 consecutive days of inactivity:
Players who only play for a short time either delete their character in the first month, or just forget about the game and are automatically deleted.
Apart from this step, the probability to leave the game slightly decreases with a player's age.

\section{Network Properties}\label{sec:defNW}
\subsection{Directed and Undirected Networks}
A network $\mathcal{G}$, called graph in mathematics, consists of a set $\mathcal{N}$ of \emph{nodes} $n_i$
and a set $\mathcal{L}$ of \emph{links} connecting these nodes: $\mathcal{G} := (\mathcal{N}, \mathcal{L})$ \cite{Wasserman1994,Dorogovtsev2003,Newman2010Book}.
In a directed network, links are ordered pairs: $\mathcal{L} \ni l_{ij} := (n_i, n_j)$
is a link from node $n_i$ to node $n_j$.
In an undirected network, links are unordered pairs of nodes: $\mathcal{L}_\mathrm{undir} \ni l_{ij} := \{n_i, n_j\}$.
Here, a node represents a player in the game, while a link represents an interaction between two players.
For each of the interaction types we studied, we produce a separate network, denoted by a superscript
on the associated quantities.
The (directed) links are constructed in the following way:
\begin{description}
 \item[$l_{ij} \in \mathcal{L}^\mathrm{trade}$] if player $i$ has traded with a building of player $j$, 
 \item[$l_{ij} \in \mathcal{L}^\mathrm{comm.}$] if a message has been sent from player $i$ to player $j$,
 \item[$l_{ij} \in \mathcal{L}^\mathrm{friend}$] if player $i$ has marked player $j$ as friend,
 \item[$l_{ij} \in \mathcal{L}^\mathrm{enemy}$] if player $i$ has marked player $j$ as enemy.
\end{description}
The \emph{symmetrization} $\mathcal{G}_{\mathrm{undir}} = (\mathcal{N}, \mathcal{L}_{\mathrm{undir}})$
of a directed network $\mathcal{G} = (\mathcal{N}, \mathcal{L})$ is constructed in the following way:
Starting with $\mathcal{G}_{\mathrm{undir}, 0} = (\mathcal{N}, \mathcal{L}_{\mathrm{undir}, 0})$, where $\mathcal{L}_{\mathrm{undir}, 0}=\emptyset$,
a link $l_{ij}$ is added to $\mathcal{L}_{\mathrm{undir}, 0}$ if $l_{ij} \in \mathcal{L}$ or if $l_{ji} \in \mathcal{L}$.

\subsection{Degree}
In an undirected network, the \emph{degree} $k_{\mathrm{undir}, i}$ of $n_i$ is the number of other nodes $n_j$,
for which the link $l_{ij}$ is present in the network, $k_{\mathrm{undir}, i} := \#\{ n_j : l_{ij} \in \mathcal{L}_\mathrm{undir} \}$
(where $\#\{\dots\}$ means the cardinality, i.e. the number of elements of the set).
$\mathcal{N}_i := \{ n_j : l_{ij} \in \mathcal{L}_\mathrm{undir} \}$ is called the set of (nearest) \emph{neighbors} of node $n_i$.
By $k_{\mathrm{nn}, i}$ we denote the mean degree of the neighbors of $n_i$, $k_{\mathrm{nn}, i} := \langle k_j \rangle_{\mathcal{N}_i}$.
In a directed network, there are two degrees: The \emph{indegree} $k_{\mathrm{in}, i}$ of node $n_i$ is the number of other nodes $n_j$,
from which a link points to node $n_i$ in the network, $k_{\mathrm{in}, i} := \#\{ n_j : l_{ji} \in \mathcal{L} \}$.
Accordingly, the \emph{outdegree} $k_{\mathrm{out}, i}$ of node $n_i$ is the number of other nodes $n_j$,
to which a link points from node $n_i$ in the network, $k_{\mathrm{out}, i} := \#\{ n_j : l_{ij} \in \mathcal{L} \}$.

\subsection{Clustering Coefficient}
In an undirected network, the \emph{clustering coefficient} $C_i$ of node $n_i$ is the ratio of pairs of neighbors of $n_i$ that are connected
to the number of all pairs of neighbors of $n_i$,
\begin{equation*}
C_i := \frac{\#\{l_{jk} \in \mathcal{L} : n_j \in \mathcal{N}_i \wedge n_k \in \mathcal{N}_i \}}{\frac 1 2 k_i(k_i-1)}.
\end{equation*}
Clustering coefficients have been calculated using the function '\texttt{clustercoeffs}' from the package 'gaimc' by David Gleich\footnote{available at \texttt{https://github.com/dgleich/gaimc}}.

\section{Linear Regression Model}

\begin{table*}[!htp]
\caption{\bf{Linear regression model for wealth.}}
 \begin{tabular}{r|*{5}{l}}
 & day 240 & day 480 & day 720 & day 960 & day 1200 \\
constant & $9.3\times 10^5$ & $-7.61\times 10^6$ & $-1.63\times 10^7$ & $-1.06\times 10^7$ & $-1.49\times 10^7$ \\
age & $-6.52\times 10^{3**}$ & $-9.73\times 10^{3***}$ & $-9.41\times 10^{3***}$ & $-1.15\times 10^{4****}$ & $-1.19\times 10^{4***}$ \\
activity $a$ & $5.57^{****}$ & $7.87^{****}$ & $9.67^{****}$ & $10.2^{****}$ & $13.1^{****}$ \\
faction rank & $5.44\times 10^{5****}$ & $3.56\times 10^{5***}$ & $4.25\times 10^{5***}$ & $6.21\times 10^{5***}$ & $6.28\times 10^{5**}$ \\
XP & $33.7^{****}$ & $12.9^{****}$ & $3.41$ & $0.139$ & $-1.49$ \\
combat skill & $-1.32\times 10^{5****}$ & $1.66\times 10^4$ & $-3.74\times 10^4$ & $-2.32\times 10^4$ & $-3.44\times 10^{5***}$ \\
farming skill & $1.47\times 10^{5**}$ & $4.4\times 10^{5****}$ & $6.44\times 10^{5****}$ & $8.47\times 10^{5****}$ & $1.68\times 10^{6****}$ \\
$f_\mathrm{trade}$ & $1.65\times 10^6$ & $1.75\times 10^6$ & $-2.4\times 10^6$ & $4.35\times 10^6$ & $6.89\times 10^6$ \\
$f_\mathrm{messages}$ & $1.27\times 10^6$ & $-4.41\times 10^5$ & $-2.96\times 10^6$ & $-4.86\times 10^6$ & $3.54\times 10^6$ \\
$f_\mathrm{attacks}$ & $-2.01\times 10^6$ & $-2.6\times 10^6$ & $8.61\times 10^6$ & $-7.59\times 10^6$ & $-2.03\times 10^7$ \\
$f_\mathrm{good}$ & $-2.67\times 10^6$ & $2.74\times 10^5$ & $1.03\times 10^7$ & $-2.9\times 10^6$ & $-9.44\times 10^6$ \\
$k_\mathrm{in}^\mathrm{trade}$ & $4.6\times 10^{4****}$ & $1.37\times 10^{5****}$ & $2.07\times 10^{5****}$ & $2.75\times 10^{5****}$ & $4.71\times 10^{5****}$ \\
$k_\mathrm{out}^\mathrm{trade}$ & $-2.49\times 10^{4****}$ & $-3.7\times 10^{4**}$ & $-5.27\times 10^{4***}$ & $-6.96\times 10^{4**}$ & $-5.59\times 10^3$ \\
$C^\mathrm{trade}$ & $-7.87\times 10^4$ & $8.44\times 10^5$ & $5.38\times 10^5$ & $-2.42\times 10^6$ & $1.79\times 10^6$ \\
$k_\mathrm{nn}^\mathrm{trade}$ & $-1.01\times 10^{4***}$ & $-1.62\times 10^{4*}$ & $-2.96\times 10^{4*}$ & $-4.18\times 10^{4*}$ & $-8\times 10^{4**}$ \\
$k_\mathrm{in}^\mathrm{comm.}$ & $8.32\times 10^{4**}$ & $9.51\times 10^{4*}$ & $7.51\times 10^4$ & $-1.76\times 10^{5*}$ & $3.23\times 10^5$ \\
$k_\mathrm{out}^\mathrm{comm.}$ & $-6.63\times 10^{4***}$ & $-9.38\times 10^{4*}$ & $-3.14\times 10^4$ & $1.66\times 10^{5*}$ & $-1.44\times 10^5$ \\
$C^\mathrm{comm.}$ & $-1.01\times 10^{6*}$ & $-7.99\times 10^5$ & $-1.89\times 10^6$ & $-1.3\times 10^6$ & $-1.34\times 10^6$ \\
$k_\mathrm{nn}^\mathrm{comm.}$ & $-3.62\times 10^3$ & $-1.38\times 10^{4*}$ & $3.05\times 10^3$ & $7.2\times 10^3$ & $-1.33\times 10^4$ \\
$k_\mathrm{in}^\mathrm{friend}$ & $-9.37\times 10^3$ & $6.91\times 10^4$ & $1.87\times 10^3$ & $1.33\times 10^5$ & $-1.12\times 10^4$ \\
$k_\mathrm{out}^\mathrm{friend}$ & $1.14\times 10^4$ & $-7.51\times 10^{4*}$ & $-1.64\times 10^{5***}$ & $-1.85\times 10^{5***}$ & $-3.16\times 10^{5****}$ \\
$C^\mathrm{friend}$ & $1.49\times 10^{6***}$ & $1.2\times 10^6$ & $4.76\times 10^5$ & $2.99\times 10^5$ & $8.69\times 10^5$ \\
$k_\mathrm{nn}^\mathrm{friend}$ & $-1.78\times 10^{4*}$ & $-1.87\times 10^4$ & $-2.14\times 10^4$ & $-1.07\times 10^4$ & $-2.01\times 10^4$ \\
$k_\mathrm{in}^\mathrm{enemy}$ & $3.04\times 10^3$ & $-6.86\times 10^3$ & $-2.59\times 10^4$ & $-8.46\times 10^3$ & $1.18\times 10^{5***}$ \\
$k_\mathrm{out}^\mathrm{enemy}$ & $-3.51\times 10^{4*}$ & $-7.31\times 10^{4**}$ & $-4.28\times 10^4$ & $-1.28\times 10^{5**}$ & $-1.76\times 10^{5**}$ \\
$C^\mathrm{enemy}$ & $5.08\times 10^5$ & $-3.84\times 10^6$ & $-4.25\times 10^6$ & $-7.47\times 10^6$ & $1.82\times 10^7$ \\
$k_\mathrm{nn}^\mathrm{enemy}$ & $4.18\times 10^{3*}$ & $4.04\times 10^3$ & $7.03\times 10^3$ & $2.45\times 10^{4**}$ & $2.91\times 10^3$ \\
 & $r^2=0.41793$ & $r^2=0.39313$ & $r^2=0.3883$ & $r^2=0.38314$ & $r^2=0.39135$ 
\end{tabular} 
\begin{flushleft}
 Data taken at days 240, 480, 720, 960, and 1200 after the beginning of the game. %
${}^*\;p-\mathrm{value}<0.05, \;{}^{**} \;p-\mathrm{value}<0.01, \;{}^{***} \;p-\mathrm{value}<0.001, \;{}^{****} \;p-\mathrm{value}<0.0001$.
\end{flushleft}
\label{tab:Regression_full}
\end{table*}

Regressions are linear regressions in a least-square sense, using \texttt{regstats(..., 'linear')} for Tab. S\ref{tab:Regression_full} from Matlab Statistical Toolbox.



